# Temperature bandgaps and engineered thermal state access in driven nanophotonic resonators

Punnag Padhy[1*], Mohammad Asif Zaman[2], Jennifer A. Dionne[1,3*]

[1]Department of Materials Science and Engineering, Stanford University, Stanford, CA 94305, U.S.A

[2]Hansen Experimental Physics Laboratory, Stanford University, Stanford, CA 94305, U.S.A

[3]Department of Radiology, Stanford University, Stanford, CA 94305, U.S.A

*Corresponding author: punnag@stanford.edu, jdionne@stanford.edu

**Abstract**

We show that strong thermo-optic feedback in nanophotonic resonators creates forbidden steady state temperatures that are unreachable under any static excitation. This temperature bandgap opens when thermo-optic feedback gain overcomes optical and thermal dissipation, splitting an otherwise continuous thermal landscape into disconnected accessible bands. We experimentally map this temperature band structure using quasi-bound states in the continuum resonances in silicon metasurfaces. Continuous wavelength scans exploit spectrally accumulated thermal energy to access the forbidden interval, reaching up to ~88°C higher temperature than static excitation, both at the same wavelength and maximum power. A characteristic three-stage temperature rise near the band edge confirms the $-1/2$ critical exponent of saddle-node bifurcations and enables quantitative extraction of band-edge wavelengths. Combining external thermal bias with optical excitation drives programmable interband transitions, enabling nearly 8.5-fold amplification of temperature rise and wavelength-selective switching between thermal states separated by just 1 nm in excitation wavelength. The bandgap is fully designable through metasurface geometry and hybrid material integration, which tune optical absorption, confinement and resonance linewidth. These results establish the temperature bandgap as a designable degree of freedom in driven nanophotonic systems, opening routes toward all-optical thermal logic and programmable photothermal control of chemical and biological processes.

When light heats matter, the resulting temperature is generally assumed to vary continuously with illumination intensity and wavelength[1-6]. Therefore, any steady-state temperature should in principle be accessible by tuning the excitation power or wavelength. This assumption underpins photothermal applications spanning catalysis and chemical synthesis[7], photothermal therapy[8], control of cellular processes[9] and thermal management[10]. However, it also imposes dynamical limitations: the thermal response is passive and featureless, offering no programmable control over which thermal states are accessible or how the system moves between them.

Resonant photonic systems with thermo-optic feedback offer a route beyond this limitation. When light drives a resonator, the optical absorption generated temperature rise shifts the resonance through the thermo-optic effect. This in turn modifies the optical absorption to establish a feedback loop between heat and light[4,11-14], generating thermo-optical nonlinearities[5,15-23]. These nonlinearities give rise to excitation trajectory-dependent thermal memories which create different outcomes at the same illumination condition, forming the basis for thermo-optical bistability and hysteresis[15-18]. These phenomena are aptly explained by assuming an instantaneous thermal response[5,16,17]. However, which temperatures are fundamentally accessible or forbidden, independent of any excitation trajectory, remain unaddressed, and the richer dynamics that emerge from the finite thermal relaxation remain unexplored.

Here we show that strong thermo-optic feedback fundamentally restructures which steady-state temperatures nanophotonic resonators can reach. When feedback coupling at the point of maximum resonance sensitivity exceeds optical and thermal dissipation, the space of accessible steady-state temperatures undergoes saddle node bifurcation[24,25] into disconnected bands separated by a forbidden interval: a temperature bandgap. This structure is a property of the static solution landscape itself, independent of any excitation trajectory or thermal history. This threshold condition is structurally analogous to band splitting in the nearly free electron model[26], where a periodic potential couples and splits the continuous free-electron energy spectrum into allowed bands separated by a forbidden gap, and to normal-mode splitting in strongly coupled cavity-emitter systems[27]. In each case, coupling overcomes dissipation to open a forbidden interval in an otherwise continuous spectrum. The resulting band structure forms the mechanistic basis for all dynamical behavior. Within this framework, bistability and hysteresis are not fundamental properties of the nonlinear response, but consequences of different excitation trajectories intersecting the static band structure at different locations, accumulating different thermal histories, and arriving at different steady states under identical excitation conditions.

The temperature bandgap is not merely a theoretical construct but a physically accessible and engineerable feature that dictates the dynamics of driven photonic systems. We show that excitation protocols such as continuous wavelength scans exploit the system's thermal memory to access the bandgap, reaching steady-states that static excitation cannot at any power or wavelength. We also show that the remnant influence of the annihilated equilibrium pair[28-31] due to the bifurcation[24,25] generates a three-step temperature rise at the band edge. This previously unreported time-domain response in chip-scale photonic systems exhibits a characteristic scaling that enables quantitative extraction of the band-edges. Together with external thermal bias, the near band edge dynamics enable programmable transitions between disconnected thermal bands, supporting thermally gated temperature amplification and wavelength-selective switching. Resonator geometry and material composition provide independent handles for bandgap engineering, establishing forbidden thermal states as a designable rather than intrinsic property.

Using amorphous silicon (aSi) metasurfaces supporting quasi-bound states in the continuum (q-BIC) resonances[32-37] as a model platform, we experimentally map the full thermal band structure, characterize its dynamical signatures, and demonstrate systematic bandgap engineering through geometry and hybrid material integration. These results establish the temperature bandgap as a new degree of freedom for driven nanophotonic systems, with direct implications for thermal logic, programmable photothermal control of chemical and biological processes, and the experimental study of nonlinear dynamics in feedback-driven systems.

## Results

### Feedback-driven restructuring of the thermal solution space

We use q-BIC metasurfaces consisting of rows of aSi nanoblocks on fused silica substrates supporting guided-mode resonances as our experimental platform (Fig. 1a&b; fabrication details in Methods)[32-37]. While nanoblock geometry dictates the resonance position, their periodic width perturbations ($\Delta d$) couple the bound modes to free-space laser illumination. The rows are spaced 1 μm apart to maximize the number of illuminated unit cells while suppressing significant inter-row modal coupling.

Resonant absorption modifies both the real and imaginary parts of the aSi refractive index[4,11-14] ($\frac{dn_{Si}}{dT} \approx 3.2 \times 10^{-4}/K$ and $\frac{d\kappa_{Si}}{dT} \approx 1 \times 10^{-5}/K$), inducing a resonance shift that feeds back into heat generation (SI Note-1). The resultant photothermal dynamics are described by the thermal balance equation[12]:

$$M_{th}\frac{\partial T}{\partial t} = P_{abs}(P_{ex}, \lambda_0 + \psi\Delta T - \lambda_{ex}) - \beta_{dis}\Delta T \quad (1)$$

where absorbed power ($P_{abs}$) follows a Lorentzian dependence on the effective detuning ($\Delta\lambda_{eff} = \lambda_0 - \lambda_{ex} + \psi\Delta T$) between the excitation wavelength ($\lambda_{ex}$) and the thermally shifted resonance ($\lambda_0 + \psi\Delta T$). Here $M_{th}$ is the thermal mass, $P_{ex}$ is the excitation power, ψ and $\Gamma_\lambda$ are the thermo-optic shift coefficient and resonance linewidth (FWHM) respectively in wavelengths and $\beta_{dis}$ is the thermal dissipation rate. Although the model is expressed in frequency detuning, wavelength detuning is used for experimental comparison (SI Note- 2 & 3).

Without thermo-optic feedback (ψ = 0), the temperature increases linearly with power and tracks the linear spectral response (Fig. 1c & d, black curves). At low powers and detunings, weak feedback distorts this linear response while maintaining a continuous, single-valued thermal landscape (Fig. 1c & d, dashed orange curves). However, at sufficiently high power and detuning, strong nonlinear feedback introduces turning points and disconnected steady-state temperature solution branches separated by forbidden bands (Fig. 1c & d, solid orange curves with grey forbidden bands).

These turning points are saddle-node bifurcations[24,25] (Fig. 1e) at which steady-state solutions annihilate in stable-unstable equilibrium pairs, opening a finite interval of temperatures that are inaccessible under static excitation at that specific detuning and any power: the temperature bandgap. The band edges are defined by the simultaneous conditions for thermal equilibrium and marginal stability:

$$\frac{\partial T}{\partial t} = 0, \frac{\partial}{\partial T}\left(\frac{\partial T}{\partial t}\right) = 0 \quad (2)$$

The first condition balances heat generation against dissipation. The second identifies vanishing restoring force when the thermo-optic feedback, set by the peak slope of the Lorentzian absorption profile, equals the dissipation rate ($\frac{\partial P_{abs}}{\partial T} = \beta_{dis}$). A temperature bandgap only emerges when the maximum feedback exceeds $\beta_{dis}$. Normalized analytical solution of both conditions (SI Note-4) shows that band edges emerge at an initial detuning $\left(\Delta\lambda_0 = \lambda_0 - \lambda_{ex}\right)$ of $-\sqrt{3}\frac{\Gamma_\lambda}{2}$ and a temperature of $\frac{2\sqrt{3}}{3}\frac{\Gamma_\lambda}{2\psi}$, placing the effective detuning precisely at the Lorentzian inflection point ($\Delta\lambda_{eff} = -\frac{1}{\sqrt{3}}\frac{\Gamma_\lambda}{2}$), the point of maximum sensitivity to thermo-optic feedback (SI Note-5). At this threshold, the feedback and dissipation pathways satisfy $\frac{\psi P_{abs}}{\frac{\Gamma_\lambda}{2}\beta_{dis}} = \frac{8\sqrt{3}}{9}$. Physically, the forbidden interval opens when the feedback overcomes both optical and thermal dissipation.

Phase-space trajectories ($\frac{\partial T}{\partial t}$ vs $T$; Fig. 1f & g) reveal how dynamics evolve with initial detuning ($\Delta\lambda_0$) and normalized excitation power ($P_{norm} = \frac{P_{ex}(\lambda_{ex})}{P_{ex,max}(\lambda_{ex})}$, SI Note-6). At initial detuning ($\Delta\lambda_0 = -3\ nm$) below the threshold ($-\sqrt{3}\frac{\Gamma_\lambda}{2} \approx -4.6\ nm$, SI Note-4), the trajectory has a single maximum and converges to a unique equilibrium (Fig. 1f) without exhibiting any band structure. Increasing detuning generates a minimum, signaling the coexistence of stable and unstable equilibria required for saddle-node bifurcations. Band splitting emerges as detuning exceeds the threshold ($\Delta\lambda_0 \leq -5\ nm$ in Fig. 1f) and the minima becomes sufficiently pronounced at $P_{norm} = 1$. At both extrema, the restoring force vanishes, critically slowing down relaxation[24,25,28,38]. The feedback gain-driven dynamical instability ($\frac{\partial}{\partial T}\left(\frac{\partial T}{\partial t}\right) > 0$) between the extrema drives rapid evolution of trajectories through this region.

At a critical detuning ($\Delta\lambda_0 = -5.95\ nm$), the trajectory minimum reaches zero, forming a saddle-node bifurcation. Its two zero-crossings define the band-edge temperatures. $P_{norm} = 1$ is the minimum power required to access the high-temperature band at this detuning. Beyond the threshold ($\Delta\lambda_0 = -6.5\ nm$), trajectories converge to a low-temperature equilibrium before reaching either extremum. Extrapolating this trajectory to incorporate the two band edges derived from steady-state analysis (Fig. 1e) depicts the three solutions of the thermal balance equation. At the band edge $\left(\Delta\lambda_0 = -5.95\ nm\right)$, the low-temperature stable equilibrium coalesces with the intermediate unstable equilibrium and annihilates, leaving only the high-temperature branch at smaller detunings; trajectories on opposite sides of the bifurcation fall into different basins of attraction, generating the abrupt interband transition that defines the forbidden region.

Near the trajectory minimum, dynamics follow the quadratic form $\frac{\partial T}{\partial t} = \mu(\Delta\lambda_0) + (T - T_{min})^2$. Here $\mu \to 0$ at the bifurcation and increases with decreasing detuning (Fig. 1f, inset). This produces a pronounced delay as trajectories linger near the "ghost" of the annihilated stable-unstable equilibrium pair[28-31] before rapidly escaping to the high-temperature branch. The ghost is therefore a dynamical remnant of the saddle-node bifurcation encoded in the curvature of the phase-space trajectory and not a mere metaphor.

At larger detuning ($\Delta\lambda_0 = -6.5\ nm$), higher power ($P_{norm} = 1.195$) is required to reach the band edges (Fig. 1g). At small detunings, the system attains a continuum of equilibrium temperatures with no bandgap (Fig. 1g, inset). Together, these results establish that strong thermo-optic feedback fundamentally restructures the accessible thermal landscape into disconnected equilibrium bands separated by a forbidden region generated through saddle-node bifurcations.

**Steady-state bandgap validation and spectral drag for forbidden state access**

Treating the forbidden temperature bands generated under static single-wavelength single-power excitation as the reference frame, we now study the trajectory-dependent dynamics and steady-state response of multiple spectral excitation protocols. The fabricated metasurfaces (Fig. 1b) support q-BIC resonances at $\lambda_0 \approx 1050\ nm$ (Fig. 2a, top panel) with quality factor $Q \approx 158$ (SI Note-7 & 8), within the tuning range of our excitation laser (SI Note-6 & 9).

Under single-wavelength excitation at $P_{norm} = 1$ (Fig. 2a, olive curve in second panel), no equilibrium temperatures appear within the theoretically predicted bandgap (SI Note-10 for experiment theory comparison). This experimentally confirms that static excitation cannot access the forbidden temperature interval.

Continuous spectral scans differ fundamentally from static excitation as the laser wavelength is swept while maintaining illumination, so the system never relaxes to its nominal resonance ($\lambda_0 \approx 1050\,nm$) between wavelength steps. The experimental forward and backward scan spectra agree with the band structure predicted by the thermal balance model (Fig. 2a). At low powers ($P_{norm} = 0.02 - 0.2$), the response shows progressive red-shifting and increasing spectral asymmetry (Fig. 2a, bottom three panels; SI Note-10), consistent with weak nonlinear feedback. At $P_{norm} = 0.6$ (Fig. 2a, third panel), the system crosses the bandgap formation threshold: the experimentally observed forward-scan red-shift of $\approx 5\,nm$ slightly exceeds the theoretical threshold of $\sqrt{3}\frac{\Gamma_\lambda}{2} \approx 4.6\,nm$, producing $\approx 1\,nm$ hysteresis width and $\approx 72$°C hysteresis depth between the forward and backward scans. The modelled bifurcation shift of $\approx 4\,nm$ falls marginally below threshold and shows no hysteresis, confirming the bandgap threshold criterion on both sides of the transition (SI Note-10). At $P_{norm} = 1$ (Fig. 2a, second panel), hysteresis widens further to $\approx 2\,nm$, with bifurcation shifts of $\approx 7\,nm$ (forward) and $\approx 5\,nm$ (backward) lying well above the threshold and agreeing with model predictions (SI Note-10). The hysteresis depth widens to $\approx 88$°C. As can be seen, it is also the peak temperature differential between the forward scan and the static single wavelength response across the entire spectrum. Crucially, the forward scan accesses temperatures within the bandgap, demonstrating spectral-drag-induced forbidden state access. This hysteretic asymmetry arises because forward and backward scans encounter the band edges in different order: the backward scan must traverse the lower band edge before reaching the upper edge. Beyond the band edge threshold, even the static excitation response must start from room temperature and traverse the lower band edge to access the high temperature branch. Hence, it overlaps with the backward scan. The spectral profile of the thermal bandgap allows the forward scan to access the upper band edge starting from room temperature without crossing the lower band edge, tracing a unique route through the thermal landscape.

The physical origin of deeper spectral drag into the bandgap with power (Fig. 2b) is captured by treating the continuous spectral scan as a sequence of transitions between the phase-space trajectories governing static excitations (Fig. 2c). At $P_{norm} = 1.5$, the system adiabatically tracks the single-wavelength trajectories up to the band edge ($\Delta\lambda_0 = -7\,nm$) owing to the thermal accumulation induced strong restoring force ($\left(\left|\frac{\partial}{\partial T}\left(\frac{\partial T}{\partial t}\right)\right| \gg 0\right)$). At this detuning, $P_{norm} = 1.5$ is the minimum power needed to access the high-temperature band. Beyond this detuning, two competing constraints confine it to the bandgap: accumulated thermal energy prevents bifurcation into the low-temperature branch, while $P_{norm} = 1.5$ is insufficient to sustain the high-temperature branch. We refer to this regime as spectral drag because the trajectories possess no corresponding static excitation solutions.

As detuning increases from $-8\,nm$ to $-11\,nm$ within this regime (Fig. 2c), the restoring force weakens until critical slowing down is reached at $\Delta\lambda_0 =- 11\,nm$. This triggers breakdown of adiabatic tracking at the next wavelength step ($\Delta\lambda_0 =- 12\,nm$) and delayed bifurcation[38] into the low-temperature branch relative to static excitation (which bifurcates at $\Delta\lambda_0 =- 8\,nm$), enabling access to the forbidden band. This critical slowing down and delayed bifurcation are confirmed experimentally and validated by the model at the maximum experimental power of $P_{norm} = 1$ (Fig. 2d and Fig. 2a, second panel from top). The observed rapid relaxation away from the bandgap erases memory of past states, thereby converging the forward scan with the static excitation response. However, the memory persists near the bandgap due to slowing down of relaxation, sustaining the forward scan's departure from the static curve.

Thermal accumulation during spectral drag enables continuous scans to reach $\approx 15°C$ higher peak temperatures than static excitations or continuous increasing power scans (Fig. 2e; SI Note-11&12), establishing spectral drag as an effective strategy for maximizing photothermal heating. Notably, the highest temperatures occur at wavelengths that yield low-temperature equilibria under static excitation, demonstrating that maximal heating requires first reaching the high-temperature branch and then dragging the system spectrally into the bandgap.

**Ghost dynamics at the band edge**

Having characterized the spectral response near the band edge, we examine the corresponding time-domain dynamics. Under static excitation far from the band edge ($\Delta\lambda_0 =- 4\,nm$, $P_{norm} = 1$), the system reaches its high-temperature steady state nearly instantaneously (Fig. 3a). Near the band edge (near $\Delta\lambda_0 =- 5\,nm$), however, the temperature response exhibits a characteristic three-stage evolution : an initial linear rise, a rapid transition to the high-temperature branch, and slow relaxation to the high-temperature steady state. This behavior stems from the ghost of the annihilated equilibrium pair (Fig. 1f): the initial slow rise reflects the trajectory approaching the now-absent low-temperature stable equilibrium, while the subsequent rapid jump arises as the trajectory escapes the coalesced repulsive unstable equilibrium.

The quadratic ghost dynamics near the band edge (Fig. 1f) yield $T \sim \sqrt{\mu}\, tan[\sqrt{\mu}(t - t_0)]$. Thus, the delay time associated with the initial linear rise scales as $t_{Delay} \propto \frac{1}{\sqrt{\mu}}$. As μ varies linearly with blue detuning from the band edge, $t_{Delay} \propto \frac{1}{\sqrt{(\Delta\lambda_0 - \Delta\lambda_{BE})}}$, in excellent agreement with experiment (Fig. 3b). This $-1/2$ exponent is a universal characteristic of saddle-node bifurcations irrespective of the system. From the fit we extract a band-edge detuning of $\Delta\lambda_{BE} =- 5.19\,nm$, a spectral parameter from time domain measurements.

The divergence of $t_{Delay}$ reflects the growing influence of ghost dynamics near the bifurcation. Upon escaping this state, the trajectory transitions abruptly to the high-temperature branch in the absence of any restoring force at the saddle-node. This inability to relax rapidly into intermediate steady states gives rise to the temperature bandgap, distinguishing its dynamical origin from conventional electronic bandgaps defined by static band structure.

These results provide direct experimental evidence for the saddle-node origin of the band edge. The observed delay-time scaling confirms that ghost dynamics governs trajectories near the bifurcation and provides a quantitative, time-domain route to estimate the band-edge parameters that define the forbidden thermal interval.

**Probing thermal bias-driven interband transitions using ghost dynamics**

Having established the steady-state band structure and the dynamical signatures of the band edge, we show that the temperature bandgap can be directly probed and navigated using external thermal bias, establishing it as a physically controllable feature and not merely a mathematically elegant representation of the observed dynamics. Analogous to electrical biasing in electronic devices, an applied thermal offset enables programmable interband transitions and an independent extraction of band-edge temperatures.

Under static excitations at $P_{norm} = 1$ (Fig. 4a, olive curve), the system bifurcates from the high to low-temperature branch beyond a critical detuning ($\Delta\lambda_0 =- 4\, nm$). Therefore, to probe bias-driven (SI Note-13) interband transitions, we monitor the band-edge dynamics in the time-domain, under combined optical and thermal excitation for $\Delta\lambda_0 \leq - 4\, nm$. Applying a thermal bias that pushes the system towards the low-temperature band edge, produces the characteristic delayed transition governed by the ghost state behavior observed in Fig. 3. The delay time indicates proximity to the band edge.

Deploying this diagnostic (Fig. 4b), we introduce an external thermal bias sufficient to drive the system from the low to the high temperature branch, under optical illumination. Abrupt interband transitions upon reaching threshold temperatures enable direct extraction of band edges. Increasing detuning from $\Delta\lambda_0 =- 5\, nm$ to $- 6\, nm$ requires progressively larger bias to trigger the transition, reflecting the detuning-dependent shift of the bifurcation boundary. Increasing the bias reduces the delay time by pushing the system towards the low-temperature band edge. The transition temperature remains largely the same for each detuning irrespective of the bias-dependent delay, validating the underlying band structure behavior. At $\Delta\lambda_0 =- 5\, nm$, the extracted low and high temperature band edges ($\approx 52°C$ and $\approx 121°C$) differ from those in Fig. 3a ($\approx 47°C$ and $\approx 108°C$) by $5°C$ and $13°C$ respectively, attributable to the elevated substrate temperature under external bias and possible device drift. This indicates cross-protocol consistency. The high temperature band edges show discrepancies of $19°C$ and $6°C$ relative to the model at $\Delta\lambda_0 =- 5\, nm$ and $- 6\, nm$ (model: $102°C$ and $139°C$), while the low temperature band edges show discrepancies of approximately $12°C$ at both detunings, as the model assumes the substrate temperature at the ambient reference. Additional uncertainty arises from camera sampling rate and estimates of thermo-optic coefficient and linewidth (SI Note-14).

Pre-applying thermal bias before optical excitation (Fig. 4c) across larger detunings ($- 7\, nm$ to $- 9\, nm$) reveals full trajectory control over disconnected thermal branches. A sufficiently large positive bias initializes the system within the high-temperature attraction basin, where optical excitation alone puts the system deep within the low temperature branch. On the other hand, a cooling bias blue-shifts the system resonance. This drives the effective detuning at $\Delta\lambda_0 =- 4\, nm$ above the bandgap threshold ($- \sqrt{3}\frac{\Gamma_\lambda}{2} \approx - 4.6\ nm$). Hence, it exhibits ghost-dynamics-mediated delayed transitions (Fig. 4c, inset) before transitioning into the high temperature branch. In both cases, transition temperatures provide the band-edge values. Mapping this extracted bandgap spectrum (Fig. 4a), we see that the static excitation temperatures do not access this range while the forward spectral scan expectedly penetrates it.

Together, these results establish bias-driven interband transitions as a quantitative route for extracting band-edge temperatures. Crucially, ghost dynamics is established as an intrinsic feature of the system response that governs its approach to the band edge regardless of the physical pathway.

**Thermal amplification and wavelength-selective switching through interband transition**

Having established the temperature bandgap as a measurable and programmable feature of the thermal landscape, we demonstrate that it enables thermal-state processing through programmable control of thermal states and trajectories.

We first demonstrate externally gated thermal amplification (Fig. 5a). Under static optical excitation at $\Delta\lambda_0 = -5\,nm$ and $P_{norm} = 1$, the system remains in the low-temperature branch, producing only a modest temperature rise ($\Delta T_{low} \approx 12°C$). An applied bias alone produces an even smaller response ($\Delta T_{bias} \approx 3.6°C$). When combined, however, the bias drives the system across the band edge into the high-temperature branch, yielding $\Delta T_{high} \approx 100°C$, nearly $8.5$-fold amplification from a small external temperature bias. This gain arises directly from proximity to the band edge, allowing a sub-threshold perturbation to trigger a large interband transition.

The relationship between amplification gain and delay time provides a quantitative characterization of this effect across operating conditions (Fig. 5b). As the applied thermoelectric voltage increases from $0.1\,V$ to $0.5\,V$, the bias temperature rise ($\Delta T_{bias}$) increases (Fig. 5b, inset). As a result, $t_{Delay}$ decreases substantially while the high-temperature steady state increases only slightly (Fig. 4b). Therefore, the bias-normalized amplification gain ($G_{norm} = \frac{\Delta T_{amp}}{\Delta T_{bias}}$) decreases with increasing bias. The delay time scales approximately as $t_{Delay} \propto \sqrt{G_{norm}}$ (SI Note-15), consistent with the $-1/2$ critical exponent of saddle-node ghost dynamics. The scaling is approximate because the slow time-varying thermal bias $\Delta T_{bias}(t)$ (SI Note-13) makes $\mu_{eff} = \mu + \psi \Delta T_{bias}(t)$ (SI Note-15) time-varying rather than a fixed control parameter. These results establish a clear operating principle: proximity to the band edge, tuned through applied bias, is the unified control parameter for thermally gated amplification. Operating closer to the edge maximizes energy-normalized gain at the cost of longer delay.

We next demonstrate wavelength-selective thermal switching arising from the detuning dependence of the band-edge threshold (Fig. 1e). Under optical excitation alone at $P_{norm} = 1$, both $\Delta\lambda_0 = -5\,nm$ and $-6\,nm$ detunings remain confined to the low-temperature branch. As the low-temperature band edge occurs at higher temperature for $-6\,nm$ than for $-5\,nm$ (Fig. 5c, dashed horizontal lines), an appropriately chosen bias selectively drives only the $-5\,nm$ trajectory across the band edge into the high-temperature state, enabling deterministic wavelength-selective switching between two channels separated by $1\,nm$ under a common bias.

Together, these results demonstrate that programmable control of band-edge proximity converts the temperature bandgap into a functional platform for nonlinear thermal-state processing. The sensitivity of ghost dynamics at the band edge provides a unified framework for thermal amplification, wavelength-selective switching, and, by extension, thermal logic operations inaccessible to conventional continuous photothermal systems.

**Engineering the thermal bandgap through metasurface geometry and hybrid material design**

Having established the nonlinear dynamical origin of the temperature bandgap and demonstrated control over interband transitions, we show that the bandgap can be systematically engineered through metasurface geometry and hybrid material integration. These approaches modify optical confinement, absorption and resonance linewidth, which together govern photothermal heating and thermo-optic feedback (SI Note-16).

Geometric control is demonstrated by varying the perturbation parameter $\Delta d$, which tunes the radiative scattering rate $\left(Q_{scat}\right)$ and thus the overall $Q$ (Fig. 6a). Reducing $\Delta d$ to $50\,nm$ suppresses scattering losses, narrows the resonance linewidth, and increases optical confinement, amplifying both photothermal heating and thermo-optic sensitivity. The stronger nonlinear feedback causes band splitting to emerge at lower initial detuning ($\Delta\lambda_0 = -\sqrt{3}\frac{\Gamma_\lambda}{2} \approx -\ 2.5\,nm$), with increased hysteresis width and depth (Fig. 6a, top panel). Conversely, increasing $\Delta d$ broadens the resonance, weakens optical confinement, and reduces the feedback gain below the dissipation rate, suppressing bandgap formation entirely. Therefore, no nonlinear signatures are observed at $\Delta d = 200\,nm$ and $300\,nm$ (Fig. 6a, bottom two panels). Metasurface geometry thus provides control over the thermal band structure by engineering optical scattering.

Material-level control is demonstrated by integrating a thin $1\,nm$ gold film onto the metasurface (Fig. 6b). Gold increases optical absorption, enhances photothermal heating and nonlinear feedback, despite reducing $Q$. The net effect is an increase in both hysteresis width and depth and deeper forward scan penetration into the bandgap compared to bare aSi metasurface at the same excitation power. We also obtain $\approx$ 30℃ rise in temperature via spectral drag compared to static excitation which is higher than the $\approx$ 13℃ rise for bare aSi metasurface. This establishes that optimal bandgap engineering requires maximizing feedback gain through heat generation and thermo-optic sensitivity while curtailing dissipation, not simply maximizing $Q$. As with bare aSi metasurfaces, the system exhibits critical slowing down of the relaxation dynamics before delayed bifurcation of the forward spectral scan into the low temperature band (Fig. 6b, inset) and ghost dynamics at the band edge (SI Note-17). Ghost dynamics enables experimental prediction of the band edges without complete knowledge of the material properties of the hybrid system (SI Note-17). The temperature-dependent optical response of gold partially opposes the silicon thermo-optic contribution[39,40], reducing the effective thermo-optic coefficient of the hybrid system. This contribution becomes more pronounced with increasing gold thickness to generate qualitatively new behavior (see SI Note-17 for the photothermal response of metasurface with $2\,nm$ thick gold coating). Hybrid material integration therefore opens routes to thermal band structures inaccessible through geometric design alone.

Together, geometry and material composition provide complementary handles on the temperature bandgap: geometry primarily controls optical confinement and feedback gain through linewidth, while material integration modifies absorption and introduces competing feedback pathways. These results establish that the temperature bandgap emerges from mesoscale engineering of the feedback between optical response and thermal generation, a flexible design framework for tailoring nonlinear thermal responses across photonic platforms and material systems.

**Discussions**

The results establish two foundational insights into the thermal physics of dielectric resonators with thermo-optic feedback. First, mapping the complete steady-state solution structure of the thermal balance equation under static single-wavelength single-power excitation, reveals the temperature bandgap as a global partition of the thermal landscape into disconnected accessible bands and a forbidden interval, independent of any excitation trajectory and thermal history. It provides the reference frame within which the steady-state and dynamical response of any excitation protocol can be interpreted. Second, treating temperature as a fully autonomous dynamical variable, by retaining the finite $\frac{\partial}{\partial T}\left(\frac{\partial T}{\partial t}\right)$ rather than assuming an instantaneous response, provides mechanistic insight into the bandgap and reveals a richer class of dynamical phenomena that steady-state analysis alone cannot yield.

Complementary to prior studies that characterized thermo-optic nonlinearity through fast thermal responses away from the band edge, the phase portrait $\frac{\partial T}{\partial t}$ vs $T$ reveals three dynamical consequences. First, the marginal stability condition ($\frac{\partial}{\partial T}\left(\frac{\partial T}{\partial t}\right) = 0$) connects the previously reported bistability emergence thresholds to the balance between the maximum thermo-optic feedback coupling at the Lorentzian inflection point and the combined optical and thermal dissipation. This coupling-induced origin of the temperature bandgap structurally unifies it with band splitting in the nearly free electron model and normal-mode splitting in strongly coupled cavity-emitter systems. The same framework also indicates that critical slowing down at the band edge is the source of the thermal memory underlying bistability and hysteresis. Second, continuous wavelength scan trajectories spectrally drag with the excitation wavelength into the forbidden interval due to two competing constraints: thermal accumulation preventing bifurcation into the low-temperature branch due to slowing relaxation at the band edge, and insufficient power preventing continuation on the high-temperature branch. However, fast relaxation away from the band edge dissipates memory to converge with the static excitation response. Together these establish where spectral scan trajectories reside relative to the static thermal landscape. Third, ghost dynamics near the band edge with its characteristic delay time provides a common dynamical framework for interband transitions, temperature gated temperature amplification, and wavelength-selective switching. The interband transition temperature remains fixed by detuning alone, independent of the applied bias, confirming it as a property of the static band structure rather than the excitation history. Together, spectral drag and ghost dynamics constitute previously unaccessed thermo-optical dynamics in nanophotonic systems that emerge from finite thermal relaxation and establish the bandgap as a dynamically rich region of their thermal landscape.

The temperature bandgap differs functionally from electronic bandgaps and conventional phase transitions[41]. Its width and position are continuously tunable through resonator geometry and material composition rather than fixed by crystal symmetry. Its dynamically unstable band edges contrast with both the static band edges of electronic systems and the intrinsic instabilities of phase transitions.

The threshold condition for bandgap emergence at the Lorentzian inflection point is material platform-agnostic. It can be satisfied by lithium niobate microresonators[42] as well as silicon-rich silicon nitride[43], and lead telluride metasurfaces[44] (Table S1). The underlying criterion of the thermo-optic feedback gain exceeding optical and thermal losses is generically met across high-Q dielectric resonators on thermally resistive substrates, suggesting spectral drag, ghost dynamics, and programmable interband transitions are general features of this platform class.

Natural extensions include fano-resonant and multimodal q-BIC architectures generating multi-band thermal landscapes. Integration with phase-transition materials could hybridize abrupt structural switching with the tunable dynamical accessibility of the temperature bandgap. Coupling to endothermic chemical reactions could enable precise optical regulation of reaction equilibrium using sharp interband transitions. Arrays of resonators existing in different temperature bands can be spatially coupled through heat diffusion to achieve thermal logic-driven spatial heating profiles.

Together these results establish the temperature bandgap as a new paradigm for thermal-state engineering in driven photonic systems: one in which accessible temperatures, dynamical pathways, and bifurcation conditions are all designable. The demonstrated ability of external thermal bias to drive programmable interband transitions establishes the temperature bandgap as an open thermodynamic interface, naturally integrable with any external thermal system such as thermoelectric elements, chemical heat sources, biological thermal processes, or substrate thermal gradients to generate photothermal dynamics beyond what optical excitation alone can produce.  It can have direct implications for thermal logic and thermodynamic computing[45-47], programmable photothermal control of chemical and biological processes, and experimental access to universal nonlinear dynamics in feedback-driven systems.

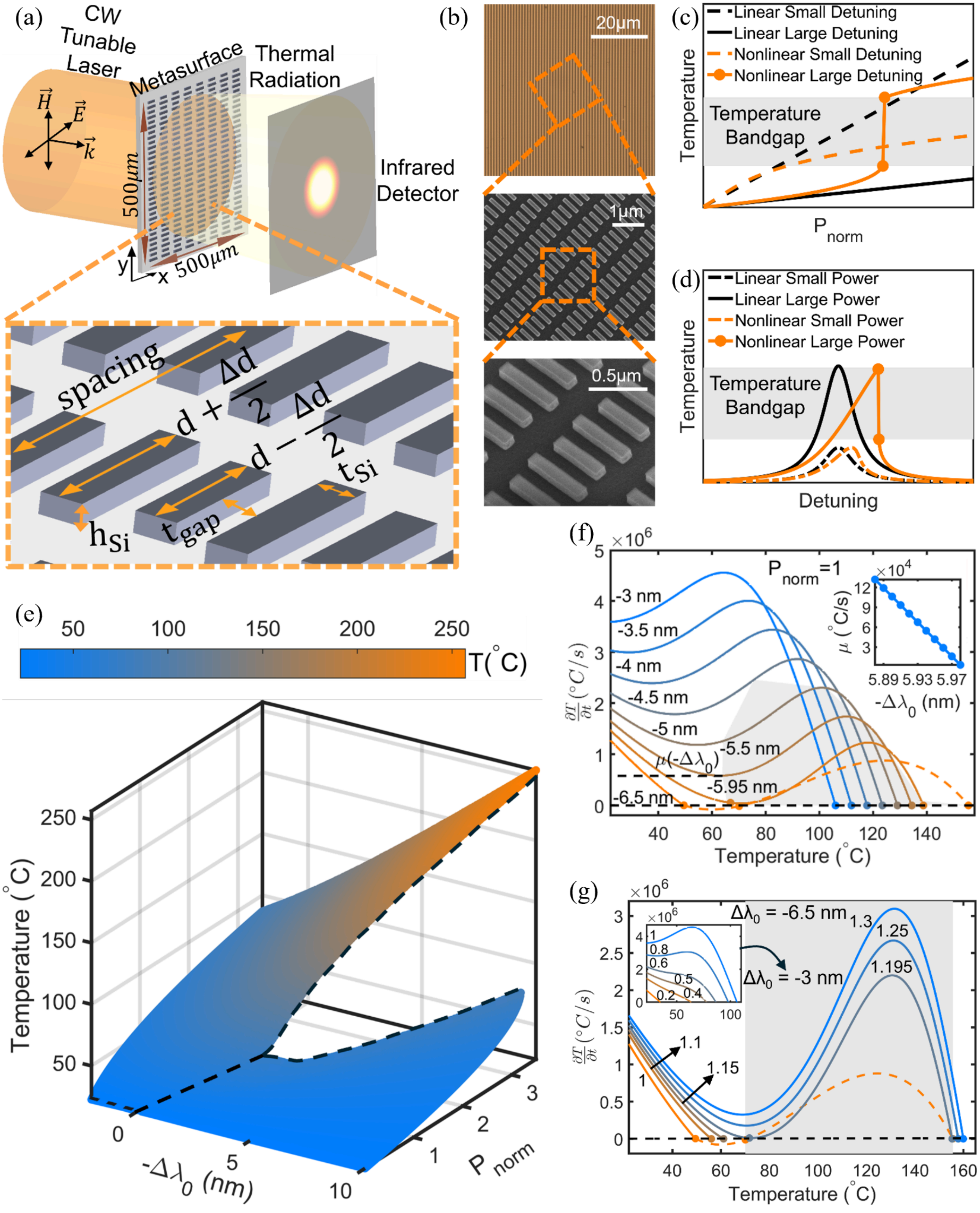

**Fig. 1. Feedback-driven restructuring of the thermal solution space and its dynamical signatures. a)** Amorphous silicon metasurface on fused silica substrate supporting quasi-bound states in the continuum (q-BIC) resonances under continuous-wave tunable laser excitation. Resonant photothermal heating is monitored through infrared thermal emission. Inset: schematic of the biperiodic metasurface geometry. **b)** Optical and scanning electron micrographs of the fabricated amorphous silicon metasurfaces comprising biperiodic nanoantenna arrays supporting guided-mode resonances. **c)** Representative linear (black), weakly nonlinear (dashed orange) and strongly nonlinear (solid orange) photothermal responses as a function of excitation power and **d)** initial detuning. Here $P_{norm} = \frac{P_{ex}(\lambda_{ex})}{P_{ex,max}(\lambda_{ex})}$. The shaded regions indicates the temperature bandgap at a specific ($\Delta\lambda_0$). **e)** Steady-state thermal landscape showing disconnected low- and high-temperature branches separated by a forbidden interval bounded by saddle-node bifurcations. $P_{norm} > 1$ is used only for modelling. **f)** Phase-space trajectories $\left(\frac{\partial T}{\partial t}\ vs\ T\right)$ as a function of initial detuning ($\Delta\lambda_0$) **g)** and normalized excitation power ($P_{norm}$). The grey region denotes the detuning-dependent temperature bandgap. Inset of **f)** depicts the evolution of the trajectory minimum (μ) with detuning and **g)** depicts the absence of a bandgap at low detuning. Panels **e–g** correspond to a reference metasurface with perturbation parameter $\Delta d = 100$nm and resonance wavelength $\lambda_0 \approx 1050$nm.

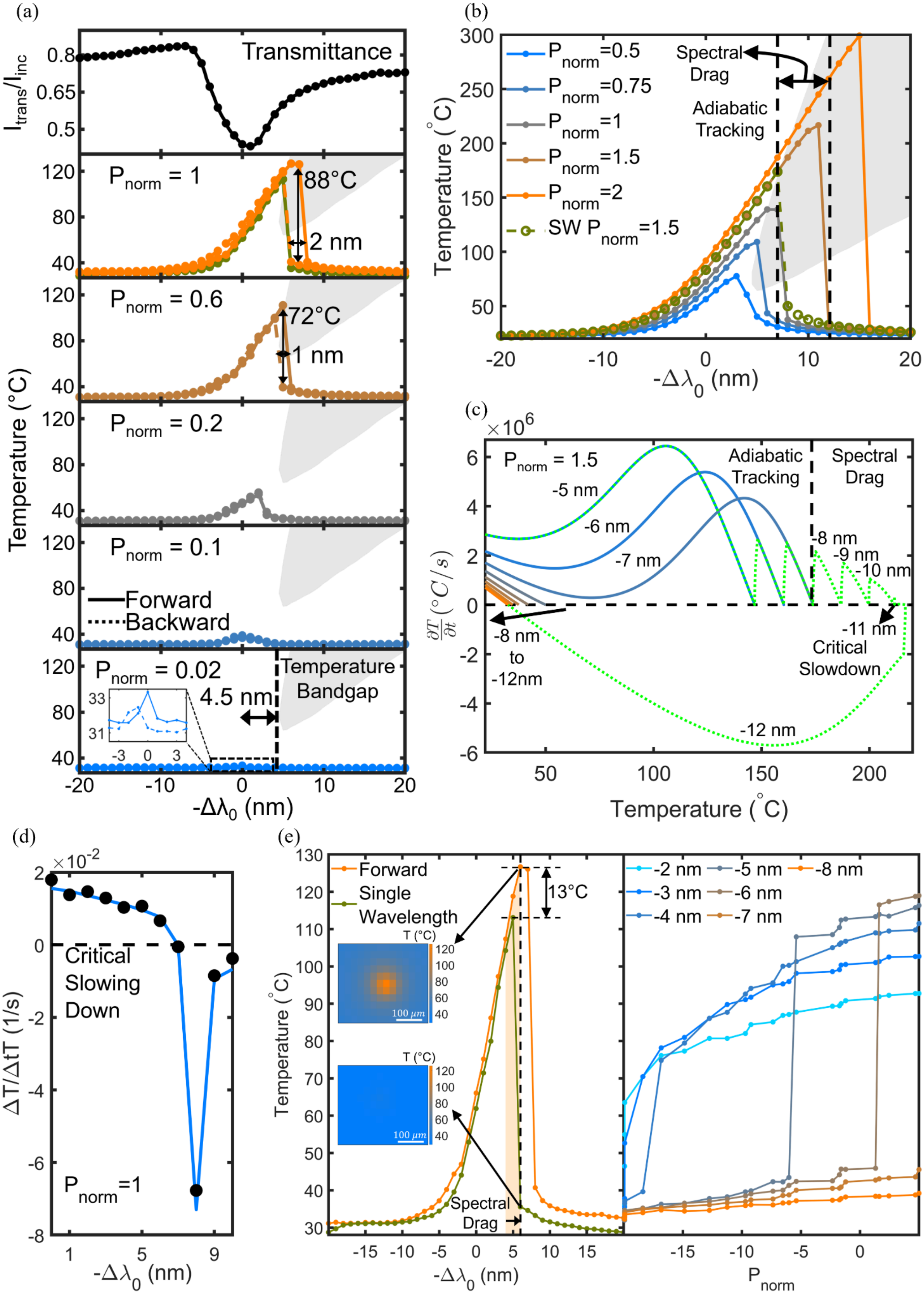

**Fig. 2. Experimental validation of temperature bandgap and spectral drag driven forbidden state access. a)** Experimental transmission spectrum of the metasurface showing a resonance centered near $\lambda_0 \approx 1050\,nm$ (top panel) and $Q \approx 158$. Lower panels: experimental photothermal response under continuous forward (solid) and backward (dashed) spectral scans with increasing normalized excitation powers ($P_{norm}$). The olive curve in the $P_{norm} = 1$ panel denotes the steady-state single-wavelength response. These spectral scans are overlayed on the theoretically predicted thermal bandgaps (grey shaded region) and suggest good quantitative agreement between experiments and model prediction. **b)** Simulated continuous forward spectral scans showing the emergence and progressive widening of spectral drag induced bandgap access with increasing normalized excitation power ($P_{norm}$) and subsequent delayed bifurcation into the low temperature branch. $P_{norm} > 1$ is used only for modelling. **c)** Simulated phase-space trajectories during continuous spectral scans at $P_{norm} = 1.5$ showing spectral drag as continued adiabatic tracking of the excitation wavelength into the bandgap region. **d)** Experimental data (black dots) showing critical slowing down of continuous spectral scan at a detuning of $-7\,nm$ before undergoing rapid transition into the low temperature band at $-8\,nm$. The experimental data shows good convergence ($R^2 = 0.9916$) with models based on the thermal balance equation (blue line). **e)** Experimental comparison of photothermal heating under continuous spectral scans and static single wavelength excitation (left panel) as well as continuous increasing power scan (right panel). Continuous spectral scans attain 13°C higher peak temperatures through spectral accumulation of thermal energy compared to the single wavelength excitation. Insets show thermal images of trajectory-dependent difference of $\approx$ 88°C in temperature between continuous spectral scan and single-wavelength excitation at the same detuning ($\Delta\lambda_0 = -6\,nm$).

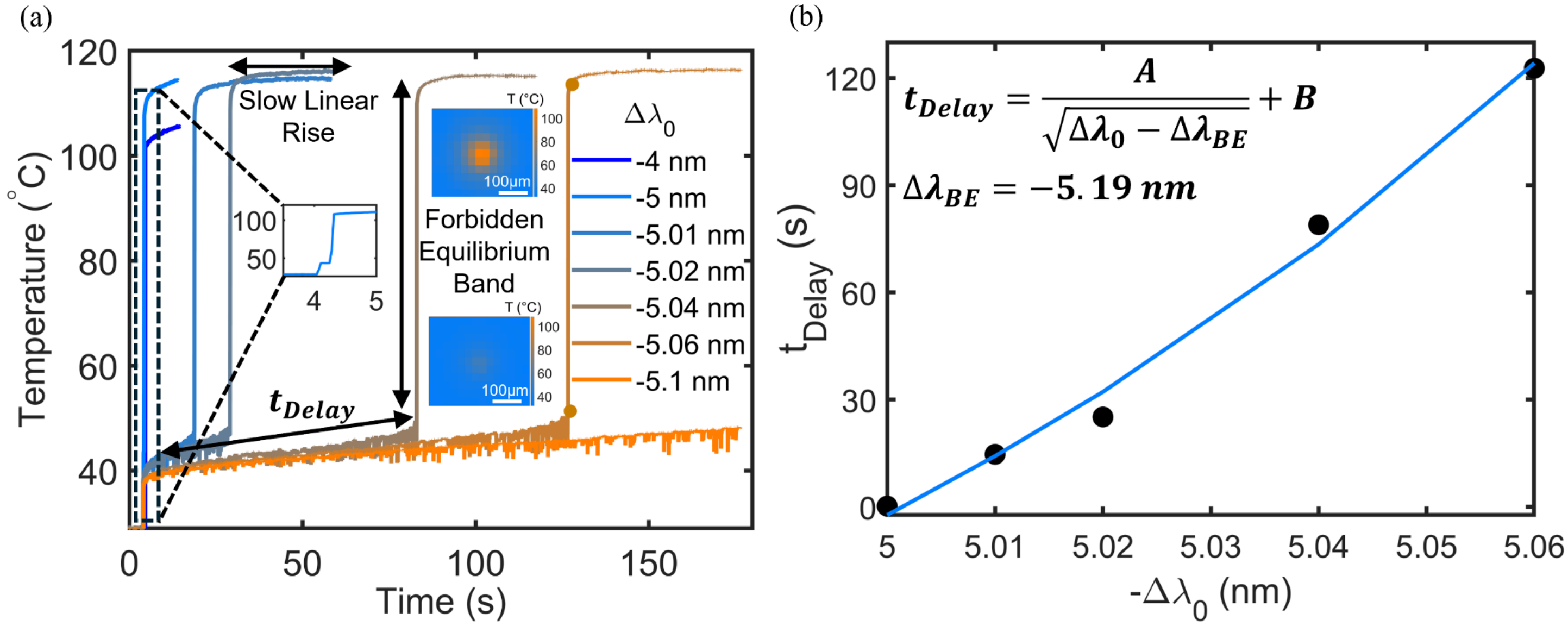

**Fig. 3. Ghost dynamics at the band edge. a)** Time-resolved photothermal response under single-wavelength excitation near the high-temperature band edge showing a characteristic three-stage evolution comprising an initial approximately linear rise which represents the delayed regime near the band edge, followed by a rapid transition to the high-temperature branch before slowly relaxing to steady state. Insets show infrared thermal images before and after the interband transition. **b)** Scaling of the delay time as a function of blue detuning from the band edge. The measured $t_{delay}$ follows the predicted dependence $t_{delay} = A/\sqrt{\Delta\lambda_0 - \Delta\lambda_{BE}} + B$, consistent with the quadratic form of the saddle-node dynamics at the minimum of the trajectory. Fitting yields a band-edge detuning $A = 255.9\ s\sqrt{nm}$, $B =- 594.3\ s$ and $\Delta\lambda_{BE} =- 5.19$ (95% CI: $[- 5.725, - 4.649]$ $nm$) with $R^2 = 0.9916$.

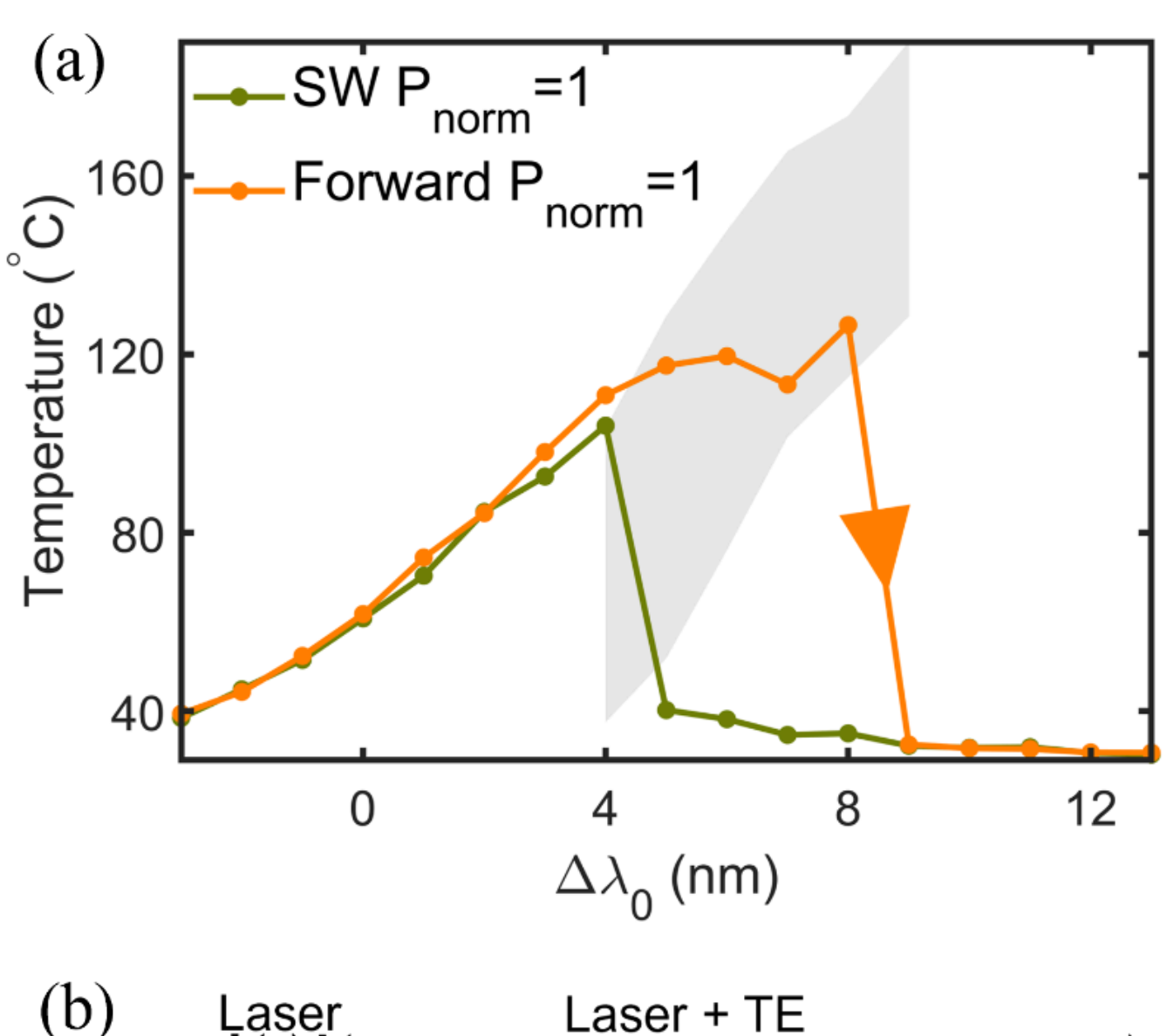

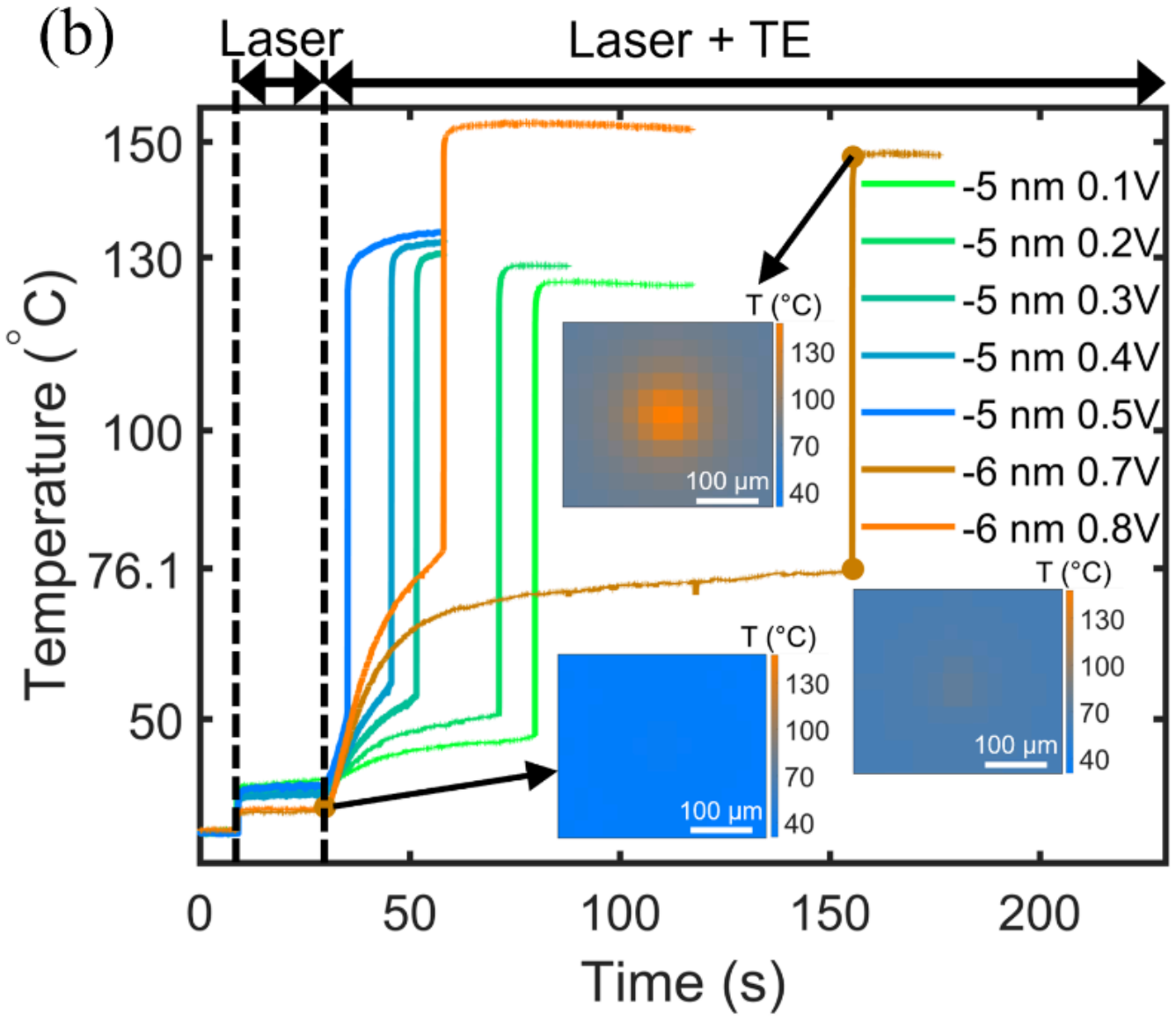

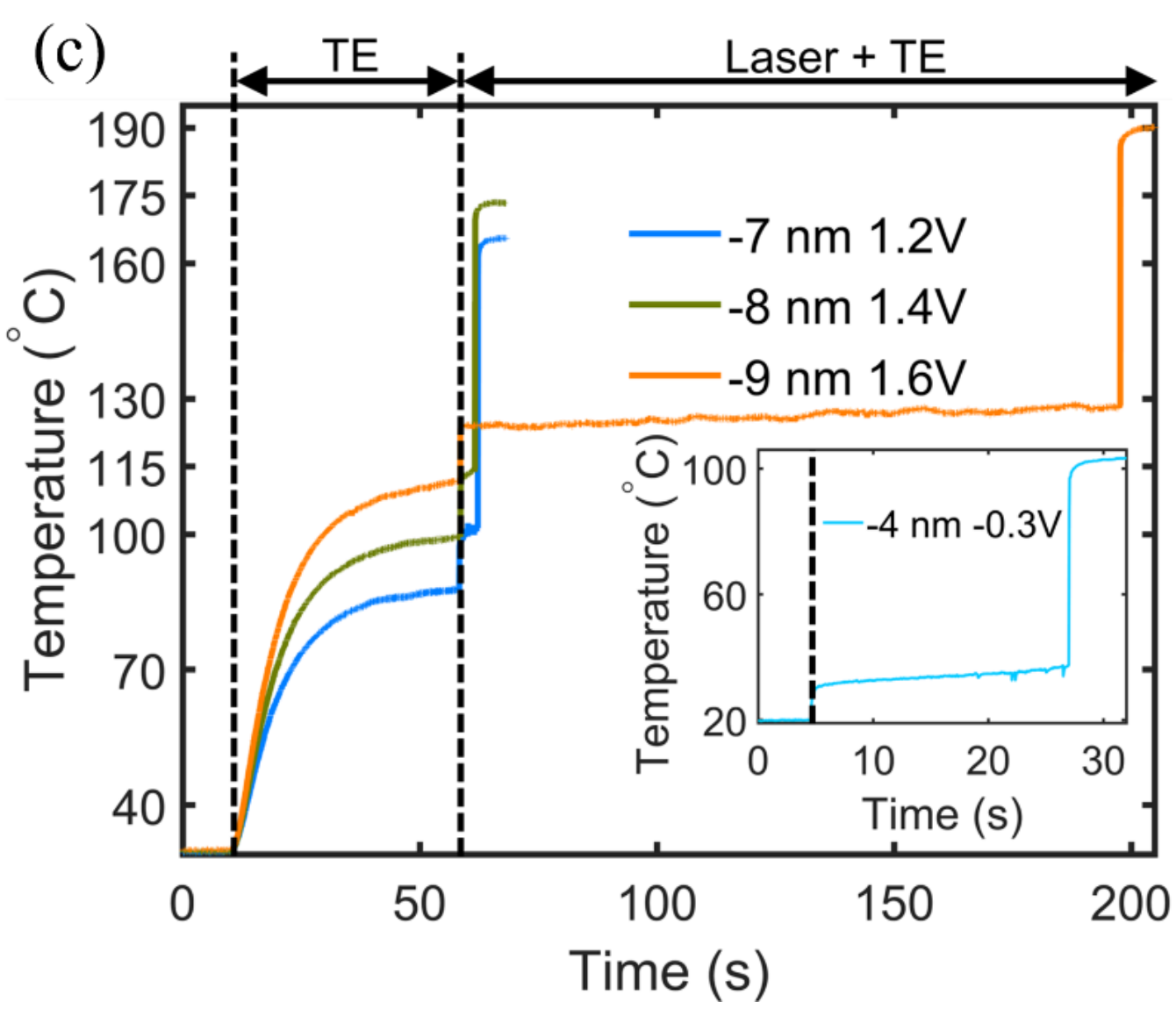

**Fig. 4. External thermal bias–driven interband transitions and extraction of temperature band edges. a)** Steady-state photothermal response under single-wavelength excitation and continuous spectral scans. The olive curve denotes the single-wavelength response at $P_{norm} = 1$, while the orange solid curve represent forward spectral scan at $P_{norm} = 1$. The grey region indicates the temperature bandgap. **b)** Bias-driven interband transitions under single-wavelength excitation at $P_{norm} = 1$ for detunings of $-5\,nm$ and $-6\,nm$ under different applied thermal biases. Insets show infrared thermal images before and after the interband transition. **c)** Time-resolved temperature dynamics under pre-applied thermal bias for detunings from $-7\,nm$ to $-9\,nm$. Inset: delayed interband transition under cooling bias at $-4\,nm$ detuning.

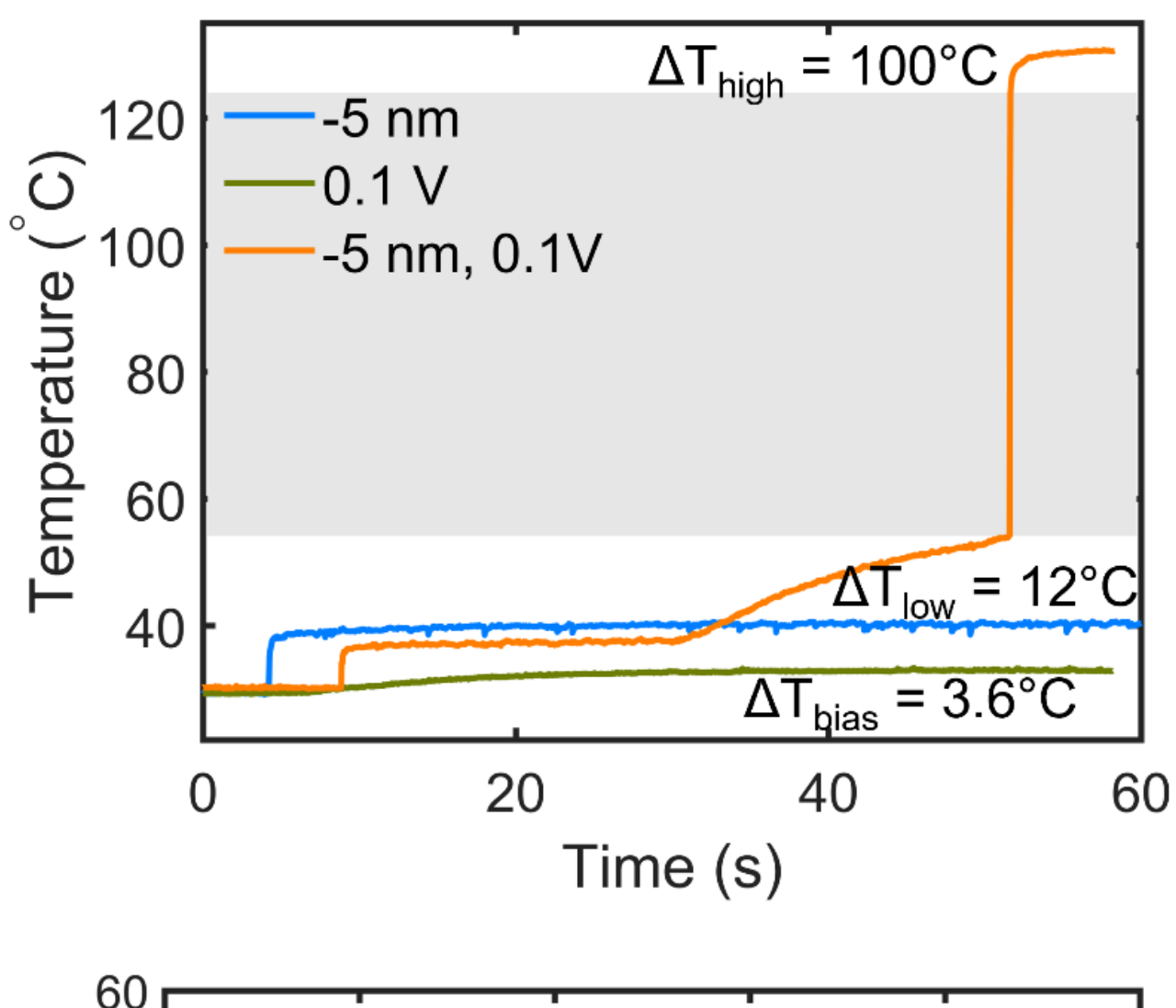

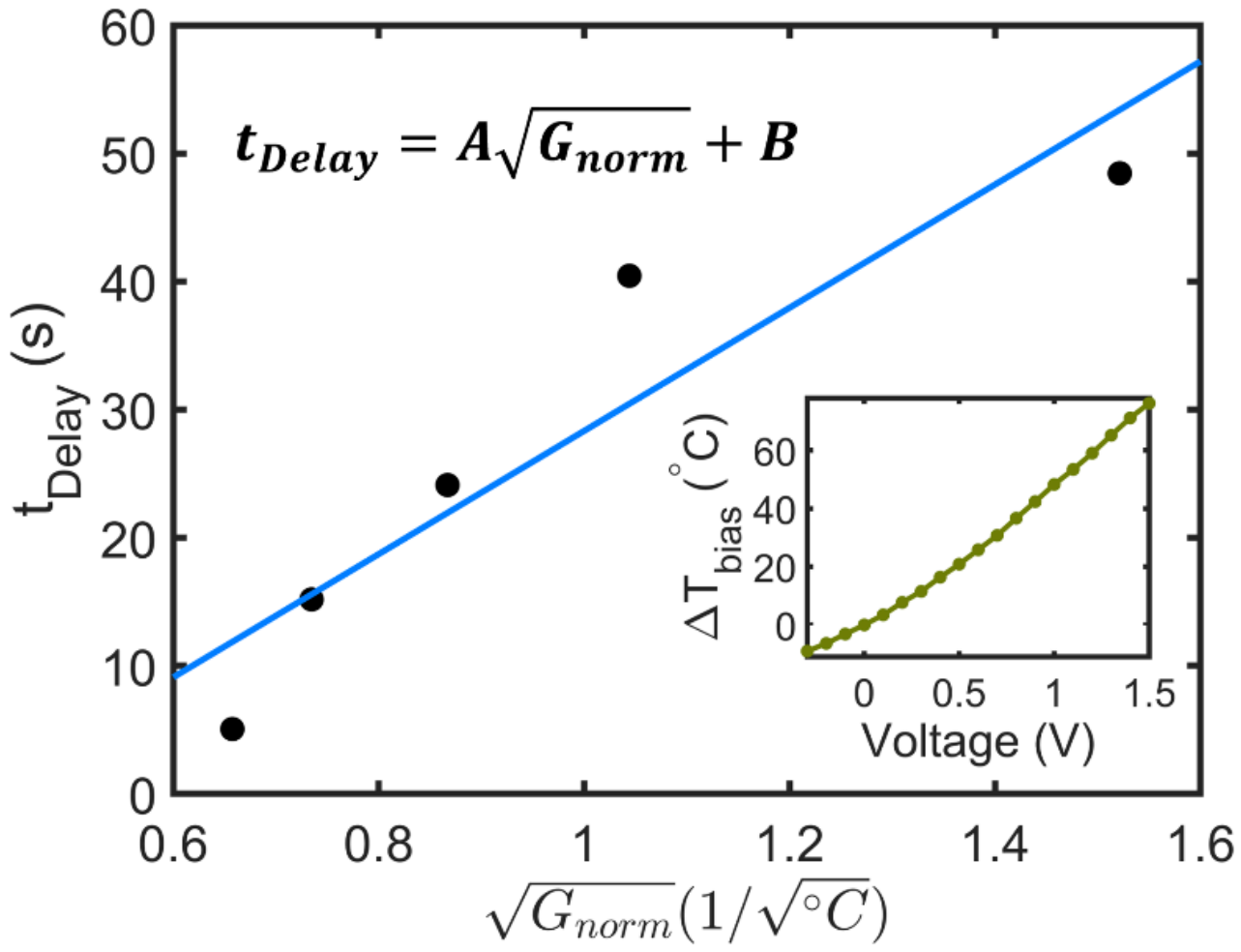

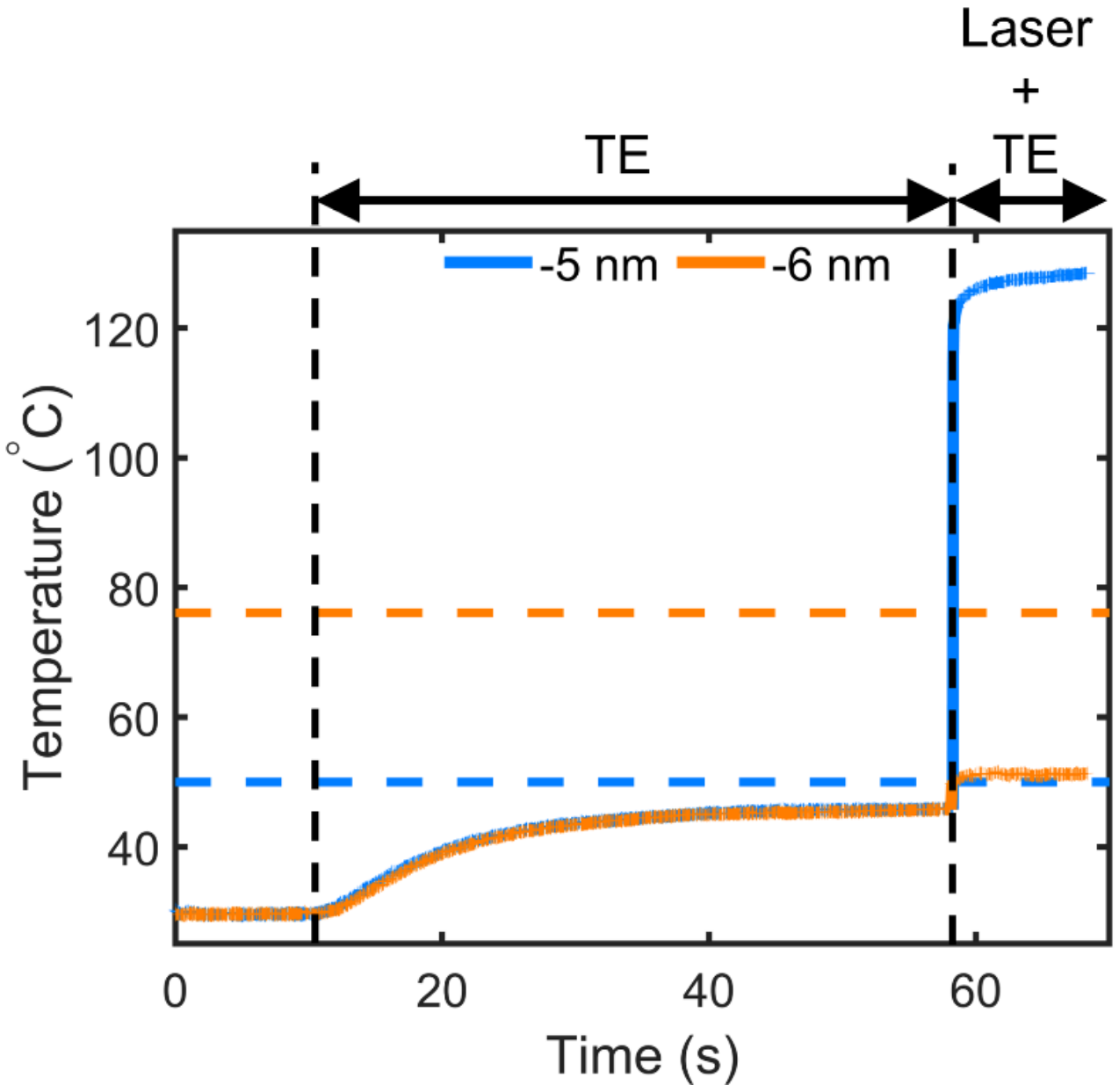

**Fig. 5. Thermal amplification and wavelength selective switching enabled by programmable thermal bias.** **a)** Externally gated thermal amplification near the band edge at $\Delta\lambda_0 = -5$nm. While the optical excitation and external thermal bias individually produce modest temperature rises of $\Delta T_{low} \approx 12°C$ and $\Delta T_{bias} \approx 3.6°C$, respectively, they combine to drive an interband transition achieving a substantially larger temperature rise of $\Delta T \approx 100°C$, yielding a $8.5$-fold amplification ($\Delta T_{amp} = \frac{\Delta T_{high}}{\Delta T_{low}}$). The shaded region denotes the temperature bandgap. **b)** Gain-delay characterization of thermally gated amplification across bias voltages from $0.1\,V$ to $0.5\,V$. The delay time between bias application and interband transition varies linearly with $\sqrt{G_{norm}} = \frac{\Delta T_{amp}}{\Delta T_{bias}}$. $R^2 = 0.86$, A = 48.1 √K·s, B = −19.8 s. The inset shows the obtained $\Delta T_{bias} \approx 3.6°C$ for varying supply voltages to the thermo-electric heating stage (SI Note-13). **c)** Wavelength-selective thermal switching enabled by detuning-dependent band-edge thresholds. Under a common applied thermal bias ($V_{bias} = 0.4$ V), the $\Delta\lambda_0 = -5$nm trajectory crosses the band edge into the high-temperature branch, whereas the $\Delta\lambda_0 = -6$nm trajectory remains confined to the low-temperature branch, yielding switching with a $1\,nm$ spectral separation. Dashed lines indicate the detuning-dependent band-edge thresholds.

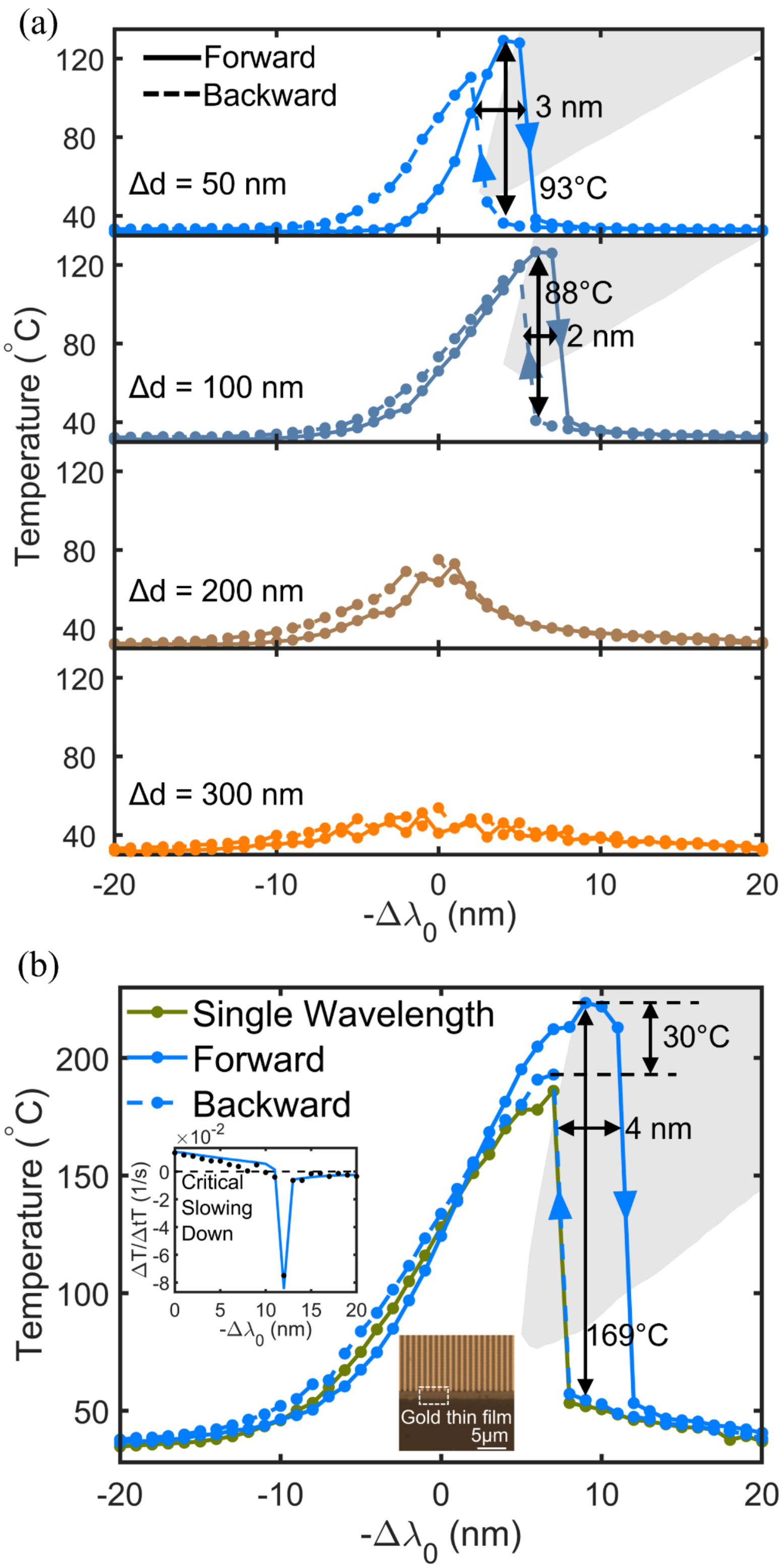

**Fig. 6. Geometric and material control of the temperature bandgap. a)** Geometric control of the temperature bandgap through variation of the perturbation parameter $\Delta d$. The reference bare amorphous silicon metasurface with $\Delta d = 100\,nm$ is shown in the second panel from top. Reduced radiative scattering by decreasing $\Delta d$ to $50\,nm$ increases $Q$ to $\approx\ 204$ which lowers the critical detuning threshold for emergence of the bandgap. By enhancing optical confinement, it increases photothermal heat generation, thermo-optic sensitivity and nonlinear feedback. Therefore, increasing $\Delta d$ to $200\, nm$ and $300\, nm$ progressively suppresses the nonlinear response and eliminates the temperature bandgap by decreasing $Q$ to $\approx 95$ and $\approx 60$ respectively. **b)** Material-level control through integration of a thin ($1\, nm$) gold film. Increased photothermal generation arising from enhanced optical absorption compensates for the reduction in $Q$ to $\approx 94$, leading to increased hysteresis width ($4\, nm$) and depth ($169°C$) together with deeper penetration of the forward spectral scan into the bandgap. Due to the higher thermal accumulation, the forward spectral scan achieves $30°C$ higher peak temperature than the static single wavelength response. Inset shows optical micrographs of the gold-coated metasurface and critical slowing down at $\Delta\lambda_0 = -11\, nm$ before delayed bifurcation into the low temperature branch for the forward spectral scan.

## Materials and Methods

### *Electromagnetic Simulations of Metasurface Resonance*

Numerical electromagnetic simulations were performed on a single unit cell using COMSOL Multiphysics 6.2. The illumination was set normal to the metasurface and periodic boundary conditions were employed in the other two directions. The real and imaginary parts of the refractive index of glass (fused silica), air and amorphous silicon (aSi) are $n_{Si} = 3.62$, $\kappa_{Si} = 0.007$, $n_{glass} = 1.445$, $n_{air} = 1$. Glass and air were assumed to be optically lossless. Eigenvalue solvers are used to obtain resonance positions and Q-factors.

### *Photothermal Simulations of Metasurface Resonance*

The resonant photothermal heating simulations were implemented in Matlab using the model represented by equation 1. The experimental or finite element simulation-based extraction of the various model parameters are provided in supplementary notes. The real and imaginary parts of the thermo-optic coefficient of amorphous silicon used in the simulations are: $\frac{\partial n_{Si}}{\partial T} = 3.2 \times 10^{-4}/°C$ and $\frac{\partial \kappa_{Si}}{\partial T} = 1 \times 10^{-5}/°C$. For the metasurfaces with $1\, nm$ film of gold, the effective thermo-optic coefficients used in the simulations were: $\frac{\partial n_{eff}}{\partial T} = 2.94 \times 10^{-4}/°C$ and $\frac{\partial \kappa_{Si}}{\partial T} = 1 \times 10^{-5}/°C$

### *Fabrication*

The metasurfaces were defined in amorphous silicon (aSi) films on fused silica substrates using standard electron-beam lithography and plasma etching. Fused silica wafers were diced into $1\, cm$×$1\, cm$ chips. After a piranha clean, $100\, nm$ of aSi was deposited on the chips using plasma enhanced chemical vapor deposition. Following a thorough rinse and a dehydration bake, the chips were spin coated with an adhesion layer (Surpass 4000), a negative resist (maN-2401) and a charge dissipation layer (eSpacer). The resists were patterned using a 100 kV Raith electron-beam system and developed using MF-319. After the patterns were transferred onto aSi films by plasma etching and stripping the resists, the chips were cleaned again in piranha. For metallization, a thin Ti/Au layer (2 Å Ti, $1\, nm$ or $2\, nm$ Au) was evaporatively deposited following standard photolithography on $1.6\, \mu m$ SPR 3612 positive photoresist. Liftoff in acetone transfers the patterns onto the metal layer.

### *Experimental Setup*

A Toptica tunable laser along with a booster is used as the source (SI Note-9). The collimated output out of the fiber passes through a linear polarizer following which a half wave plate is used to rotate its polarization to line up the direction of the incident electric field along the length of the silicon nanobars. A 125 mm biconvex lens focusses the incident light on the back focal plane of the 20x objective, with 0.45 numerical aperture (N.A.) and a working distance between $8.18\, mm$ - $8.93\, mm$, after passing through a beam splitter. The objective then illuminates the metasurface mounted on a x-y translation stage with a near parallel beam of light. The transmitted beam is filtered out by a long pass filter with a pass band from $7.6\, \mu m$ - $14.6\, \mu m$. Only the thermal radiation within this wavelength range reaches the lens of the infrared/thermal camera which then focuses it on the detector to provide the studied thermal information. The reflected beam is blocked off by a beam block.

### *Experimental Process*

The continuous spectral scans are carried out by illuminating the metasurface at the starting wavelength of either the forward or backward scan, letting the temperature to stabilize for $10\, s$ and then increasing/decreasing the wavelength by $1\, nm$ while keeping the illumination on. For the static excitations, after ensuring that the metasurface is at room temperature, the illumination is switched on for $10\, s$ to let the temperature stabilize before blocking it off. After estimating the high temperature to low temperature bifurcation point, a series of experiments are carried out over a narrow range of wavelengths around it to extract the linear rise time, the critical temperature and the high temperature steady state. The thermal camera captures temperatures with a spatial resolution of $\approx 25.64\ \mu m$ (SI Note-18) and with a time resolution of $33\ ms$. For the experiments involving measurements with external thermal bias, the metasurface chips are mounted on a custom-made thermoelectric stage (SI Note-13). After letting the applied thermal bias to stabilize for $50\ s$, the laser illumination is switched on. For the reverse order of activation, the thermal biases are switched on after the laser illumination has been switched on for $10\ s$.

**Acknowledgements:** We appreciate Pumpkinseed for support with the tunable laser setup, Chih-Yi Chen for help with SEM imaging and Varun Dolia, Remi Dado, Amy Mckeown-Green and Priyanuj Bordoloi for feedback on the manuscript. P.P. and J.A.D acknowledge funding support from the Gordon and Betty Moore Foundation for capital equipment, AFOSR MURI for device fabrication, characterization and salary. P.P. acknowledges Stanford OTL HIT fund for a portion of the salary. Part of the work was performed at nano@stanford (RRID: SCR_026695)

**Author Contributions:** P.P. and J.A.D. conceived the idea. P.P. fabricated the devices, did the experiments and the numerical modelling. P.P. and M.A.Z. designed the thermoelectric sample stage for temperature bias. M.A.Z made the stage. J.A.D. supervised the project.

**Competing Interest Statement:** The authors declare no competing interests.

**Data, Materials and Software Availability:** All studied data are included in the article or Supplementary Information.

## References

1. X. Cui, Q. Ruan, X. Zhuo, X. Xia, J. Hu, R. Fu, Y. Li, J. Wang, H. Xu, Photothermal nanomaterials: a powerful light-to-heat converter. *Chem. Rev.* **123**, 11 6891-6952 (2023).
2. E. Cortes, F. J. Wendisch, L. Sortino, A. Mancini, S. Ezendam, S. Saris, L. de S. Menezes, A. Tittl, H. Ren, S. Maier, Optical metasurfaces for energy conversion. *Chem. Rev.* **122**. 19, 15082-15176 (2022).
3. P. K. Jain, X. Huang, I. H. El-Sayed, M. A. El-Sayed, Noble metals on the nanoscale: optical and photothermal properties and some applications in imaging, sensing, biology and medicine. *Acc. Chem. Res.* **41**, 12, 1578-1586 (2008).
4. G. P. Zograf, M. I. Petrov, S. V. Makarov, Y. S. Kivshar, All-dielectric thermonanophotonics. *Adv. Opt. Photonics* **13**, 3, 643-702 (2021).
5. D. Ryabov, O. Pashina, G. Zograf, S. Makarov, M. Petrov, Nonlinear optical heating of all-dielectric super-cavity: efficient light-to-heat conversion through giant thermorefractive bistability. *Nanophotonics*, **11**, 17 3981-3991 (2022).
6. G. Baffou, R. Quidant, Nanoplasmonics for chemistry, *Chem. Soc. Rev*. **43**, 11, 3898-3907 (2014).
7. L. Yuan, Y. Zhao, A. Toma, V. Aglieri, B. Gerislioglu, Y. Yuan, M. Lou, A. Ogundare, A. Alabastri, P. Nordlander, N. J. Halas, A quasi-bound states in the continuum dielectric metasurface-based antenna-reactor photocatalyst. *Nano Lett.* **24**, 1, 172-179 (2024).
8. H. S. Jung, P. Verwilst, A. Sharma, J. Shin, J. L. Sessler, J. S. Kim, Organic molecule-based photothermal agents: an expanding photothermal therapy universe. *Chem. Soc. Rev.* **47**, 7, 2280-2297 (2018).
9. Y. Wang, R. Garg, D. Cohen-Karni, T. Cohen-Karni, Neural modulation with photothermally active nanomaterials. *Nat. Rev. Bioeng.* **1**, 3, 193-207 (2023).
10. X. Meng, X. Wang, K. Yin, Y. Jing, L. Gu, Z. Tao, X. Ren, M. Tang, X. Shao, L. Sun, Y. Sun, Y. Dai, Y. Xiong, Integration of photothermal water evaporation with photocatalytic microplastics upcycling via nanofluidic thermal management. *Proc. Nat. Acad. Sci.* **121**, 13, e2317192121 (2024).
11. W. Koechner, Solid-state laser engineering. *Springer*, New York, U.S.A (2006).
12. R. W. Eason, A. Miller, Nonlinear optics in signal processing. Chapman and Hall, London, U.K. (1993).
13. G. Coppola, L. Sirleto, I, Rendina, M. Iodice, Advance in thermo-optical switches: principles, materials, design and device structure. *Opt. Eng.* **50**, 7, 071112 (2011).
14. F. G. Della Corte, M. Esposito Montefusco, L. Moretti, I. Rendina, G. Cocorullo, Temperature dependence analysis of the thermo-optic effect in silicon by single and double oscillator models. *J. Appl. Phys.* **88**, 12, 7115-7119 (2000).
15. K. Nishida, P. Tseng, Y. Chen, P. Wu, C. Yang, J. Yang, W. Chen, O. Pashina, M. Petrov, K. Chen, S. Chu, Optical bistability in nanosilicon with record low q-factor. *Nano Lett.* **23**, 24, 11727-11733 (2023).
16. A. Barulin, O. Pashina, D. Riabov, O. Sergaeva, Z. Sadrieva, A. Shcherbakov, V. Rutckaia, J. Schilling, A. Bogdanov, I. Sinev, A. Chernov, M. Petrov, Thermo-optical bistability enabled by bound states in the continuum in silicon metasurfaces. *Laser Photonics Rev.* **18**, 2301399 (2024).
17. M. Cotrufo, A. Cordaro, D. L. Sounas, A. Polman, A. Alú, Passive bias-free non-reciprocal metasurfaces based on thermally nonlinear quasi-bound states in the continuum. *Nat. Photon.* **18**, 81-90 (2024).
18. O. C. Karaman, H. Li, E. N. Dayi, C. Galland, G. Tagliabue, Photo-thermally tunable photon-pair generation in dielectric metasurfaces. *ACS Nano* **20**, 5, 4079-4087 (2026).
19. Y. Duh, Y. Nagasaki, Y. Tang, P. Wu, H. Cheng, T. Yen, H. Ding, K. Nishida, I. Hotta, J. Yang, Y. Lo, K. Chen, K. Fujita, C. Chang, K. Lin, J. Takahara, S. Chu, Giant photothermal nonlinearity in a single silicon nanostructure. *Nat. Comm.* **11**, 4101 (2020).

20. C. Li, Y. Tang, J. Takahara, S. Chu, Nonlinear heating and scattering in a single crystalline silicon nanostructure. *J. Chem. Phys.* **155**, 20, 204202 (2021).
21. T. Zhang, Y. Che, K. Chen, J. Xu, Y. Xu, T. Wen, G. Lu, X. Liu, B. Wang, X. Xu, Y. Duh, Y. Tang, J. Han, Y. Cao, B. Guan, S. Chu, X. Li, Anapole mediated giant photothermal nonlinearity in nanostructured silicon. *Nat. Comm.* **11**, 3027 (2020).
22. Y. Che, T. Zhang, T. Shi, Z. Deng, Y. Cao, B. Guan, X. Li, Ultrasensitive photothermal switching with resonant silicon metasurfaces at visible bands. *Nano Lett.* **24**, 2, 576-583 (2024).
23. O. C. Karaman, G. N. Naidu, A. R. Bowman, E. N. Dayi, G. Tagliabue, Decoupling optical and thermal dynamics in dielectric metasurfaces for self-encoded photonic control. *Laser Photonics Rev.* **19**, 24, e01014 (2025).
24. S. H. Strogatz, Nonlinear dynamics and chaos: with applications to physics, biology, chemistry and engineering. *CRC Press*, Boca Raton, U.S.A (2024).
25. Y. A. Kuznetsov, Elements of applied bifurcation theory, applied mathematical sciences. *Springer*, New York, U.S.A (2004).
26. S. H. Simon, The Oxford Solid State Basics, Oxford University Press, Oxford, U.K. (2013).
27. M. Fox, Quantum Optics An Introduction, Oxford University Press, Oxford, U.K. (2006).
28. J. R. Tredicce, G. L. Lippi, P. Mandel, B. Charasse, A. Chevalier, B. Picque, Critical slowing down at a bifurcation. *Am. J. Phys.* **72**, 6, 799-809 (2004).
29. Z. Zheng, P. Beck, T. Yang, O. Ashtari, J. P. Parker, T. M. Schneider, Ghost states underlying spatial and temporal patterns: How nonexistent invariant solutions control nonlinear dynamics. *Phys Rev. E.* **112**, 2, 024212 (2025).
30. E. Fontich, J. Sardanyés, General scaling law in the saddle-node bifurcation: a complex phase space study. *J. Phys. A. Math Theor.* **41**, 1, 015102 (2008).
31. R. M. de Boer, C. Toebes, J. Klärs, S. R. K. Rodriguez, Ghost states of light, arXiv, https://arxiv.org/abs/2605.13584v1 (2026).
32. M. Lawrence, D. R. Barton III, J. Dixon, J. H. Song, J. van de Groep, M. L. Brongersma, J. A. Dionne, High quality factor phase gradient metasurfaces. *Nat. Nanotech.* **15**, 956-961 (2020).
33. J. Hu, F. Safir, K. Chang, S. Dagli, H. B. Balch, J. M. Abendroth, J. Dixon, P. Moradifar, V. Dolia, M. K. Sahoo, B. A. Pinsky, S. S. Jeffrey, M. Lawrence, J. A. Dionne. Rapid genetic screening with high quality factor metasurfaces. *Nat. Comm.* **14**, 4486 (2023).
34. V. Dolia, H. B. Balch, S. Dagli, S. Abdollahramezani, H. C. Delgado, P. Moradifar, K. Chang, A. Stiber, F. Safir, M. Lawrence, J. Hu, J. A. Dionne. Very-large-scale-integrated high-quality factor nanoantenna pixels. *Nat. Nanotech.* **19**, 1290-1298 (2024).
35. S. Dagli, J. Shim, H. C. Delgado, H. B. Balch, S. Abdollahramezani, C. Chen, V. Dolia, E. Klopfer, J. Dixon, J. Hu, B. Ogunlade, J. Song, M. L. Brongersma, D. Barton, J. A. Dionne, GHz-speed wavefron shaping metasurface modulators enabled by resonant electro-optic nanoantennas. *Adv. Mater.* **37**, 40, e06790 (2025).
36. F. Pan, X. Li, A. C. Johnson, S. Dhuey, A. Saunders, M. Hu, J. P. Dixon, S. Dagli, S. Lau, T. Weng, C. Chen, J. Zeng, R. Apte, T. F. Heinz, F. Liu, Z. Deng, J. A. Dionne, Room-temperature valley-selective emission in Si-$MoSe_2$ heterostructures enabled by high-quality-factor chiroptical cavities. *Nat. Comm.* **17**, 20 (2026).
37. E. Klopfer, H. C. Delgado, S. Dagli, M. Lawrence, J. A. Dionne, A thermally controlled high-Q metasurface lens. *Appl. Phys. Lett.* **122**, 22, 221701 (2023).

38. M. Fucho-Rius, S. Maretvadakethope, À. Haro, T. Alarcón, J. Sardanyés, R. Pérez-Carrasco, Local nearby bifurcations lead to synergies in critical slowing down: the case of mushroom bifurcations. *Phys. Rev. E.* **111**, 2, 024213 (2025).
39. H. Reddy, U. Guler, A. V. Kildishev, A. Boltasseva, V. M. Shalaev, "Temperature-dependent optical properties of gold thin films," *Opt. Mater. Exp.* **6**, 9, 2776–2802 (2016).
40. P. T. Shen, Y. Sivan, C. W. Lin, H. L. Liu, C. W. Chang, S. W. Chu, "Temperature- and roughness-dependent permittivity of annealed/unannealed gold films," *Opt. Exp.* **24**, 17, 19254–19263 (2016).
41. N. Goldenfeld, Lectures on phase transitions and the renormalization group, 1st edition. CRC Press, Boca Raton, U.S.A. (1992).
42. J. Wang, B. Zhu, Z. Hao, F. Bo, X. Wang, F. Gao, Y. Li, G. Zhang, J. Xu, Thermo-optic effects in on-chip lithium niobate microdisk resonators. *Opt. Exp.* **24**, 19, 21869-21879 (2016).
43. O. Goldberg, R. Gherabli, J. Engelberg, J. Nijem, N. Mazurski, U. Levy, Silicon rich nitride huygens metasurfaces in the visible regime. *Adv. Opt. Mater.* **12**, 4, 2301612 (2024).
44. T. Lewi, H. A. Evans, N. A. Butakov, J. A. Schuller, Ultrawide thermo-optic tuning of PbTe meta-atoms. *Nano Lett.* **17**, 6, 3940-3945 (2017).
45. D. Melanson, M. Abu Khater, M. Aifer, K. Donatella, M. H. Gordon, T. Ahle, Thomas G. Crooks, A. J. Martinez, F. Sbahi, P. J. Coles, Thermodynamic computing system for AI applications. *Nat. Comm.* **16**, 1, 3757 (2025).
46. L. Wang, B. Li, Thermal logic gates: computation with phonons. *Phys. Rev. Lett.* **99**, 17, 177208 (2007).
47. L. Wang, B. Li, Thermal Memory: A Storage of Phononic Information. *Phys. Rev. Lett.* **101**, 26, 267203 (2008).

# Supplementary Information:

## Temperature bandgaps and engineered thermal state access in driven nanophotonic resonators

**SI Note-1: Schematic representation of the role of thermo-optic feedback in shaping light-matter interaction and the resultant photothermal heating.**

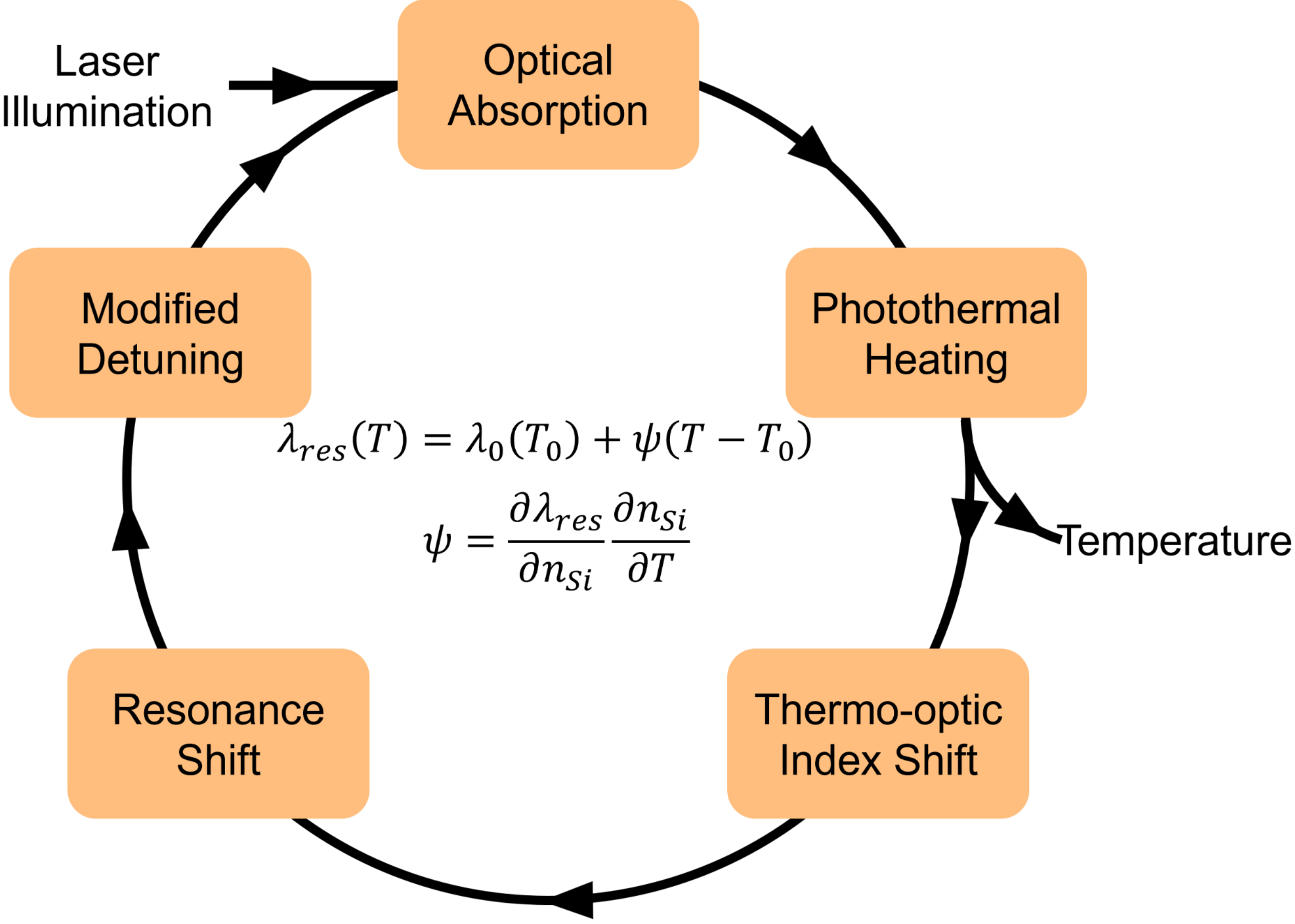

**Fig. S1. Thermo-optic feedback loop coupling optical absorption, photothermal heating, resonance shift and detuning-dependent absorption.**

**SI Note-2: Deriving thermal balance equation in terms of system parameters for numerical modelling**

The thermal balance equation (Eq. 1 in main text) in the frequency domain can be written as:

$$M_{th}\frac{\partial T}{\partial t} = P_{ex}\eta_0 L\left(\omega_0 - \omega_{ex} + \alpha\Delta T, \frac{\Gamma_\omega(\Delta T)}{2}\right) - \beta_{dis}\Delta T \quad \text{(S1)}$$

Here, $M_{th}$ is the thermal mass, $T$ is the temperature rise, $\Delta T$ is the temperature change, $P_{ex}$ is the excitation power, $\eta_0$ is the maximum absorption on resonance and $\beta_{dis}$ is the heat dissipation rate as defined in the main text. $L$ is the Lorentzian with the detuning given by $\omega_0 - \omega_{ex} + \alpha\Delta T$ and the half linewidth given by $\frac{\Gamma_\omega(\Delta T)}{2}$. $\omega_{0/ex} = \frac{2\pi c}{\lambda_{0/ex}}$, and $\alpha$ is the thermo-optic shift coefficient in the frequency domain. Knowing $\frac{\partial T}{\partial t} = \frac{\partial \Delta T}{\partial t}$ and replacing the explicit form of the Lorentzian, Eq. S1 can be written as:

$$M_{th}\frac{\partial \Delta T}{\partial t} = P_{ex}\eta_0\frac{\left(\frac{\Gamma_\omega(\Delta T)}{2}\right)^2}{\left(\omega_0 - \omega_{ex} + \alpha\Delta T\right)^2 + \left(\frac{\Gamma_\omega(\Delta T)}{2}\right)^2} - \beta_{dis}\Delta T \quad \text{(S2)}$$

As silicon has a positive thermos-optic coefficient, its refractive index will increase with temperature. As a result, its resonance will redshift, thereby making $\alpha < 0$. Using this and $\delta_0 = \omega_0 - \omega_{ex}$, Eq. S2 can be written as:

$$M_{th}\frac{\partial \Delta T}{\partial t} = P_{ex}\eta_0\frac{\left(\frac{\Gamma_\omega(\Delta T)}{2}\right)^2}{\left(\delta_0 - |\alpha|\Delta T\right)^2 + \left(\frac{\Gamma_\omega(\Delta T)}{2}\right)^2} - \beta_{dis}\Delta T \quad \text{(S3)}$$

Dividing Eq. S3 by $M_{th}$ and using $\beta_{eff} = \frac{\beta_{dis}}{M_{th}}$ we obtain:

$$\frac{\partial \Delta T}{\partial t} = \frac{P_{ex}}{M_{th}}\eta_0\frac{\left(\frac{\Gamma_\omega(\Delta T)}{2}\right)^2}{\left(\delta_0 - |\alpha|\Delta T\right)^2 + \left(\frac{\Gamma_\omega(\Delta T)}{2}\right)^2} - \beta_{eff}\Delta T \quad \text{(S4)}$$

The thermal mass of amorphous silicon ($M_{th}$) can be expressed as the product of the mass of the illuminated silicon nanostructures ($m_{Si}$) and their specific heat capacity ($C_{p,Si}$) i.e. $M_{th} = m_{Si}C_{p,Si}$. The mass ($m_{Si}$) itself can be expressed as the density of amorphous silicon ($\rho_{Si}$) times the total volume of the illuminated silicon nanostructures which is the volume of a single unit cell of the metasurface ($V_{unit-cell}$) times the number of illuminated unit cells ($N$) i.e. $m_{Si} = \rho_{Si}NV_{unit-cell}$. The number of illuminated unit cells ($N$) can be expressed as the area of the illuminating beam ($A_{beam}$) divided by the area of a unit cell ($A_{unit-cell}$) i.e. $N = \frac{A_{beam}}{A_{unit-cell}}$. Using these expressions $\frac{P_{ex}}{M_{th}}$ in Eq. S4 can be written as:

$$\frac{P_{ex}}{M_{th}} = \frac{P_{ex}}{\rho_{Si}\frac{A_{beam}}{A_{unit-cell}}V_{unit-cell}C_{p,Si}} \quad \text{(S5)}$$

Writing $\frac{P_{ex}}{A_{beam}} = I_{ex}$, Eq. S5 can be rephrased as:

$$\frac{P_{ex}}{M_{th}} = \frac{I_{ex}}{\rho_{Si} C_{p,Si} \frac{V_{unit-cell}}{A_{unit-cell}}} \tag{S6}$$

As we can see, $I_{ex}$ is the known property of the illuminating laser beam, $\rho_{Si}$ and $C_{p,Si}$ are known material properties of amorphous silicon and $\frac{V_{unit-cell}}{A_{unit-cell}}$ is a known geometrical property of the fabricated nanostructures of the metasurface. Writing $\gamma = \frac{1}{\rho_{Si} C_{p,Si} \frac{V_{unit-cell}}{A_{unit-cell}}}$, Replacing Eq. S6 in S4 we obtain:

$$\frac{\partial \Delta T}{\partial t} = \gamma I_{ex} \eta_0 \frac{\left(\frac{\Gamma_\omega(\Delta T)}{2}\right)^2}{\left(\delta_0 - |\alpha|\Delta T\right)^2 + \left(\frac{\Gamma_\omega(\Delta T)}{2}\right)^2} - \beta_{eff} \Delta T \tag{S7}$$

From Fig. 1a we can write, $V_{unit-cell} = 2dt_{Si}h_{Si}$ and $A_{unit-cell} = 2\left(t_{Si} + t_{gap}\right) \times spacing$. $\rho_{Si} = 2290 \frac{kg}{m^3}$, $C_{p.Si} = 700 \frac{J}{kg \times K}$, $\omega_{res} = \omega_0 - |\alpha|\Delta T$, $\alpha = \frac{\partial \omega_{res}}{\partial n_{Si}} \frac{\partial n_{Si}}{\partial T}$, $\Gamma(\Delta T) = \frac{\omega_0 - |\alpha|\Delta T}{Q(\Delta T)} = \frac{\omega_{res}}{Q(\Delta T)}$. Despite the metasurface exhibiting a Fano type resonance, symmetric Lorentzian profile has been used for photothermal heat generation for simplicity and due to the absence of any noticeable Fano specific feature in the experimental data. It is to be noted here that the beam spot size ($D_{beam} \approx 100\ \mu m$) is much smaller than the metasurface ($500\ \mu m \times 500\ \mu m$). The absorption at the beam edges is expected to be different from the center which is likely to exhibit an absorption efficiency close to theoretical value. To account for this an additional experimental factor ($f_e$) is introduced into the Eq. S7 as:

$$\frac{\partial \Delta T}{\partial t} = f_e \gamma I_{ex} \eta_0 \frac{\left(\frac{\Gamma_\omega(\Delta T)}{2}\right)^2}{\left(\delta_0 - |\alpha|\Delta T\right)^2 + \left(\frac{\Gamma_\omega(\Delta T)}{2}\right)^2} - \beta_{eff} \Delta T \tag{S8}$$

For a metasurface with $d = 550\ nm$, $t_{Si} = 140\ nm$, $h_{Si} = 140\ nm$ and $spacing = 1\ \mu m$, $\gamma = 22.68 \frac{Km^2}{J}$. For this bare aSi metasurface with $\Delta d = 100$, we assume $\eta_0 = 0.5$. The other relevant parameters required for the resonant photothermal modeling of the metasurface temperature are derived in SI Note-19-22. With these parameters $f_e = 0.0165$ is appropriate for experimental convergence. The model here follows a lumped element assumption, where the entire metasurface is assumed to have the same temperature and no spatial variation, for simplified analysis. Here we have assumed the peak temperature in the spatial distribution as the representative of the lumped element temperature. This is obviously higher than the mean temperature of the metasurface. So, for the bare aSi metasurface with $\Delta d = 100\ nm$, $Q \approx 198$ is used in the model, corresponding to $\kappa_{Si} = 0.007$ (SI Note-21), to ensure fit with experimental data (Fig. 2d), while the experimentally measured $Q \approx 158$ (SI Note-7). To maintain consistency, for the metasurfaces with different scattering by varying $\Delta d$, $Q$ corresponding to $\kappa_{Si} = 0.007$ are extracted. The $\eta_0$ values for each $\Delta d$ are considered as per SI Note-16. For the metasurface with $1\ nm$ gold film, $\eta_0 = 0.8113$ is used for experimental fit.

**SI Note-3: Correspondence between Lorentzian in the frequency and wavelength domain**

Using $\Delta\omega_{eff} = \omega_0 - \omega_{ex} + \alpha\Delta T$ in the Lorentzian term of Eq. S2 and then normalizing the entire term with $\left(\frac{\Gamma_\omega(\Delta T)}{2}\right)^2$, we obtain:

$$L = \frac{1}{1+\left(\frac{\Delta\omega_{eff}}{\frac{\Gamma_\omega(\Delta T)}{2}}\right)^2} \quad \text{(S9)}$$

Knowing that $\frac{d\omega}{d\lambda} = -\frac{2\pi c}{\lambda^2}$, small changes in frequency with temperature ($\Delta\omega(\Delta T)$) can be written in terms of corresponding wavelength changes ($\Delta\lambda(\Delta T)$) around the nominal resonance ($\lambda_0$) as:

$$\Delta\omega(\Delta T) = -\frac{2\pi c}{\lambda_0^2}\Delta\lambda(\Delta T) \quad \text{(S10)}$$

By writing $\Delta\omega(\Delta T)$ and $\Delta\lambda(\Delta T)$ in terms of the respective thermo-optic coefficients ($\alpha$ and $\psi$) respectively. Eq. S10 yields the following relation:

$$\alpha = -\frac{2\pi c}{\lambda_0^2}\psi \quad \text{(S11)}$$

The linewidths in the frequency and wavelength domain are also connected by the identical relation i.e.

$$\Gamma_\omega(\Delta T) = -\frac{2\pi c}{\lambda_0^2}\Gamma_\lambda(\Delta T) \quad \text{(S12)}$$

Using Eq. S11 to rewrite $\Delta\omega_{eff}$ in terms of wavelengths we obtain:

$$\Delta\omega_{eff} = 2\pi c\left(\frac{\lambda_0 - \psi\Delta T}{\lambda_0^2} - \frac{1}{\lambda_{ex}}\right) \quad \text{(S13)}$$

Replacing $\lambda_{ex} = \lambda_0 - \Delta\lambda_0$, and rearranging the terms we obtain:

$$\Delta\omega_{eff} = \frac{2\pi c}{\lambda_0^2(\lambda_0 - \Delta\lambda_0)}\left(\Delta\lambda_0\psi\Delta T - \lambda_0\Delta\lambda_0 - \lambda_0\psi\Delta T\right) \quad \text{(S14)}$$

Making the realistic assumption that $\Delta\lambda_0 \ll \lambda_0$ and $\psi\Delta T \ll \lambda_0$, we can safely say that $\Delta\lambda_0\psi\Delta T \ll \lambda_0\Delta\lambda_0$ and $\Delta\lambda_0\psi\Delta T \ll \lambda_0\psi\Delta T$. Therefore, $\Delta\omega_{eff}$ in Eq. S14 can be approximated as:

$$\Delta\omega_{eff} \approx -\frac{2\pi c}{\lambda_0^2}\left(\Delta\lambda_0 + \psi\Delta T\right) \quad \text{(S15)}$$

Using Eq. S15 and S12 in S9 and writing $\Delta\lambda_{eff} = \Delta\lambda_0 + \psi\Delta T$, we obtain:

$$L \approx \frac{1}{1+\left(\frac{2\Delta\lambda_{eff}}{\Gamma_\lambda(\Delta T)}\right)^2} \quad \text{(S16)}$$

Eq. S16 is the form used in the main text.

**SI Note-4: Solution of normalized form of thermal balance equation to identify the bandgap thresholds**

Simplifying Eq. S2 to assume time independent resonance linewidth ($\Gamma_{\omega}$), we obtain:

$$M_{th}\frac{\partial \Delta T}{\partial t} = P_{ex}\eta_0 \frac{\left(\frac{\Gamma_{\omega}}{2}\right)^2}{\left(\omega_0 - \omega_{ex} + \alpha\Delta T\right)^2 + \left(\frac{\Gamma_{\omega}}{2}\right)^2} - \beta_{dis}\Delta T \tag{S17}$$

Using $\delta_{0n} = \frac{\omega_0 - \omega_{ex}}{\frac{\Gamma_{\omega}}{2}}$, $\Delta T_n = \frac{\frac{\Gamma_{\omega}}{2}}{|\alpha|}$, $\theta = \frac{\Delta T}{\Delta T_n}$, $P_n = \frac{P_{ex}\eta_0}{M_{th}\Delta T_n}$ and $\beta_n = \frac{\beta_{dis}}{M_{th}}$, Eq. S17 can be written in the normalized form as:

$$\frac{\partial \theta}{\partial t} = \frac{P_n}{\left(\delta_{0n} - \theta\right)^2 + 1} - \beta_n \theta \tag{S18}$$

Setting $\frac{\partial \theta}{\partial t} = 0$ for the steady state condition (Eq. 2 in the main text) and using $P_N = \frac{P_n}{\beta_n} = \frac{P_{ex}\eta_0}{\beta_{dis}\Delta T_n}$, we obtain:

$$\frac{P_N}{\left(\delta_{0n} - \theta\right)^2 + 1} - \theta = 0 \tag{S19}$$

Taking the derivative of Eq. S18 with respect to θ we obtain:

$$\frac{\partial}{\partial \theta}\left(\frac{\partial \theta}{\partial t}\right) = \frac{2P_n\left(\delta_{0n} - \theta\right)}{\left[\left(\delta_{0n} - \theta\right)^2 + 1\right]^2} - \beta_n \tag{S20}$$

Setting $\frac{\partial}{\partial \theta}\left(\frac{\partial \theta}{\partial t}\right) = 0$ to satisfy the marginal stability condition (Eq. 2 in the main text), we obtain:

$$\frac{2P_N\left(\delta_{0n} - \theta\right)}{\left[\left(\delta_{0n} - \theta\right)^2 + 1\right]^2} - 1 = 0 \tag{S21}$$

Combining Eq, S19 and S21 and rearranging we obtain:

$$3\theta^2 - 4\delta_{0n}\theta + \left(\delta_{0n}^2 + 1\right) = 0 \tag{S22}$$

Solving Eq. S22 for θ in terms of $\delta_{0n}$ we obtain:

$$\theta = \frac{2\delta_{0n} \pm \sqrt{\delta_{0n}^2 - 3}}{3} \tag{S23}$$

By satisfying the steady state and the marginal stability criterion, the normalized temperature (θ) given by Eq. S23 yields the band edges. For the temperatures to be real $\delta_{0n} \geq 3$. So, the threshold normalized detuning for the emergence of the bandgap is given by:

$$\delta_{0n,th} = \sqrt{3} \tag{S24}$$

The corresponding normalized threshold temperature is:

$$\theta_{th} = \frac{2\delta_{0n,th}}{3} = \frac{2\sqrt{3}}{3} \tag{S25}$$

Using Eq. S24 and S25 in S19 we obtain the normalized threshold power as:

$$P_{N,th} = \theta_{th}\left[\left(\delta_{0n,th} - \theta_{th}\right)^2 + 1\right] = \frac{8\sqrt{3}}{9} \quad \text{(S26)}$$

More importantly replacing Eq. S24 and S25 in the normalized form of the Lorentzian in Eq. S19 we obtain:

$$L_{n,th} = \frac{1}{\left(\delta_{0n,th} - \theta_{th}\right)^2 + 1} = \frac{1}{\frac{1}{3} + 1} \quad \text{(S27)}$$

Now a normalized detuning squared of $\frac{1}{3}$ corresponds to the point of maximum slope of the normalized Lorentzian (SI Note-5). Physically this implies the point of maximum sensitivity of the optical resonance to the temperature feedback. Therefore, bandgap emerges when the maximum rate of increase in heat generation with temperature exceeds the heat loss rate. While the criteria in Eq. S24 and S26 have been reported in existing literature on thermo-optical bistability, Eq. S25 and S27 are reported in this manuscript. The criterion in Eq. S26 can be used to establish if temperature band splitting can be achieved for a nonlinear photothermal system at accessible laser powers as is shown in the table below:

| **Platform** | Reference | $\lambda_0\ (nm)$ | $\omega_0\ (THz)$ | $Q$ | $\Gamma = \frac{\omega_0}{Q}\ (THz)$ | $\beta_{eff}\ (1/s)$ | $M_{th} (J/K)$ | $\beta_{dis}\ (mW/K)$ | $\alpha\ (THz/K)$ | $P_{abs} = \frac{P_{th}\beta_{dis}\Gamma}{2\lvert\alpha\rvert}\ (mW)$ |
|---|---|---|---|---|---|---|---|---|---|---|
| Amorphous silicon metasurface | This manuscript | 1050 | 1794 | 158 | 11.35 | 73360 | $3.29 \times 10^{-10}$ | 0.024 | 0.093 | 2.25 |
| Lithium niobate microdisk resonator | S1 | 1550 | 1215 | $8.9 \times 10^5$ | $1.3 \times 10^{-3}$ | | | 5.9 | 0.018 | 0.328 |
| Silicon rich silicon nitride metasurface | S2,S3,S4 | 780 | 2415 | 500 | 4.829 | 733600 | $3.17 \times 10^{-10}$ | 0.23 | 0.049 | 17.45 |

**Table S1. List of platforms that can satisfy the threshold condition for the emergence of temperature bandgap at reasonably accessible laser powers.**

The effective thermal loss rate ($\beta_{eff}$) of the silicon rich silicon nitride (SRN) on glass is assumed to be an order of magnitude higher due to the higher thermal conductivity of SRN compared to aSi. Apart from the materials presented in the table, lead telluride metasurfaces are emerging as excellent high-Q metasurfaces in the mid-infrared regime with large thermo-optic coefficients that exhibit temperature dependent resonance shifts larger than their linewidths[S5]. This makes them excellent candidates for temperature bandgap generation.

**SI Note-5: Deriving maximum slope of the normalized Lorentzian and connecting it to bandgap threshold.**

We write the normalized Lorentzian as:

$$L_n(\Delta) = \frac{1}{\Delta^2+1} \tag{S28}$$

Taking the derivative of $L_n(\Delta)$ we obtain:

$$\frac{dL_n}{d\Delta} = -\frac{2\Delta}{\left(1+\Delta^2\right)^2} \tag{S29}$$

Its second order derivative can then be written as:

$$\frac{d}{d\Delta}\left(\frac{dL_n}{d\Delta}\right) = 2\frac{\left(3\Delta^2-1\right)}{\left(1+\Delta^2\right)^3} \tag{S30}$$

The maximum of $\left|\frac{dL_n}{d\Delta}\right|$ (symmetric) is obtained by setting $\frac{d}{d\Delta}\left(\frac{dL_n}{d\Delta}\right) = 0$. This yields:

$$\Delta^2 = \frac{1}{3} \tag{S31}$$

Eq. S31 yields the same normalized detuning for the maximum slope of the normalized Lorentzian that is represented in Eq. S27. Reorganizing Eq. S17 we obtain:

$$\frac{\partial \Delta T}{\partial t} = \frac{P_{ex}\eta_0}{M_{th}}\frac{1}{\Delta^2+1} - \frac{\beta_{dis}}{M_{th}}\Delta T \tag{S32}$$

Here, $\Delta = \delta_{0n} - \frac{\alpha}{\frac{\Gamma_\omega}{2}}\Delta T$. Taking the derivative of Eq. S32 with respect to $\Delta T$ we obtain:

$$\frac{\partial}{\partial \Delta T}\left(\frac{\partial \Delta T}{\partial t}\right) = \frac{P_{ex}\eta_0\alpha}{M_{th}\frac{\Gamma_\omega}{2}}\frac{2\Delta}{\left(1+\Delta^2\right)^2} - \frac{\beta_{dis}}{M_{th}} \tag{S33}$$

For $\frac{\partial}{\partial \Delta T}\left(\frac{\partial \Delta T}{\partial t}\right)$ to be 0 at the minimum illumination power, $\frac{2\Delta}{\left(1+\Delta^2\right)^2}$ should assume its maximum value. From Eq. S31, we know it occurs at $\Delta = \frac{1}{\sqrt{3}}$. Using it in Eq. S32 we obtain:

$$\frac{9}{8\sqrt{3}}\frac{P_{ex}\eta_0\alpha}{M_{th}\frac{\Gamma_\omega}{2}} - \frac{\beta_{dis}}{M_{th}} = 0 \tag{S34}$$

Rearranging Eq. S34 and using $P_{abs} = P_{ex}\eta_0$ we obtain:

$$\frac{P_{abs}\alpha}{\beta_{dis}\frac{\Gamma_\omega}{2}} = \frac{8\sqrt{3}}{9} \tag{S35}$$

As we can see, Eq. S35 reproduces the same threshold condition as Eq. S26. As we can see the feedback gain ( $P_{abs}\alpha$) must exceed the combined optical and thermal losses ($\beta_{dis}\frac{\Gamma_\omega}{2}$) by a certain threshold ($\frac{8\sqrt{3}}{9}$) for the emergence of bandgap-like behavior.

**SI Note-6: Power spectrum of the tunable laser with a polynomial fit for numerical modelling of resonant photothermal heating**.

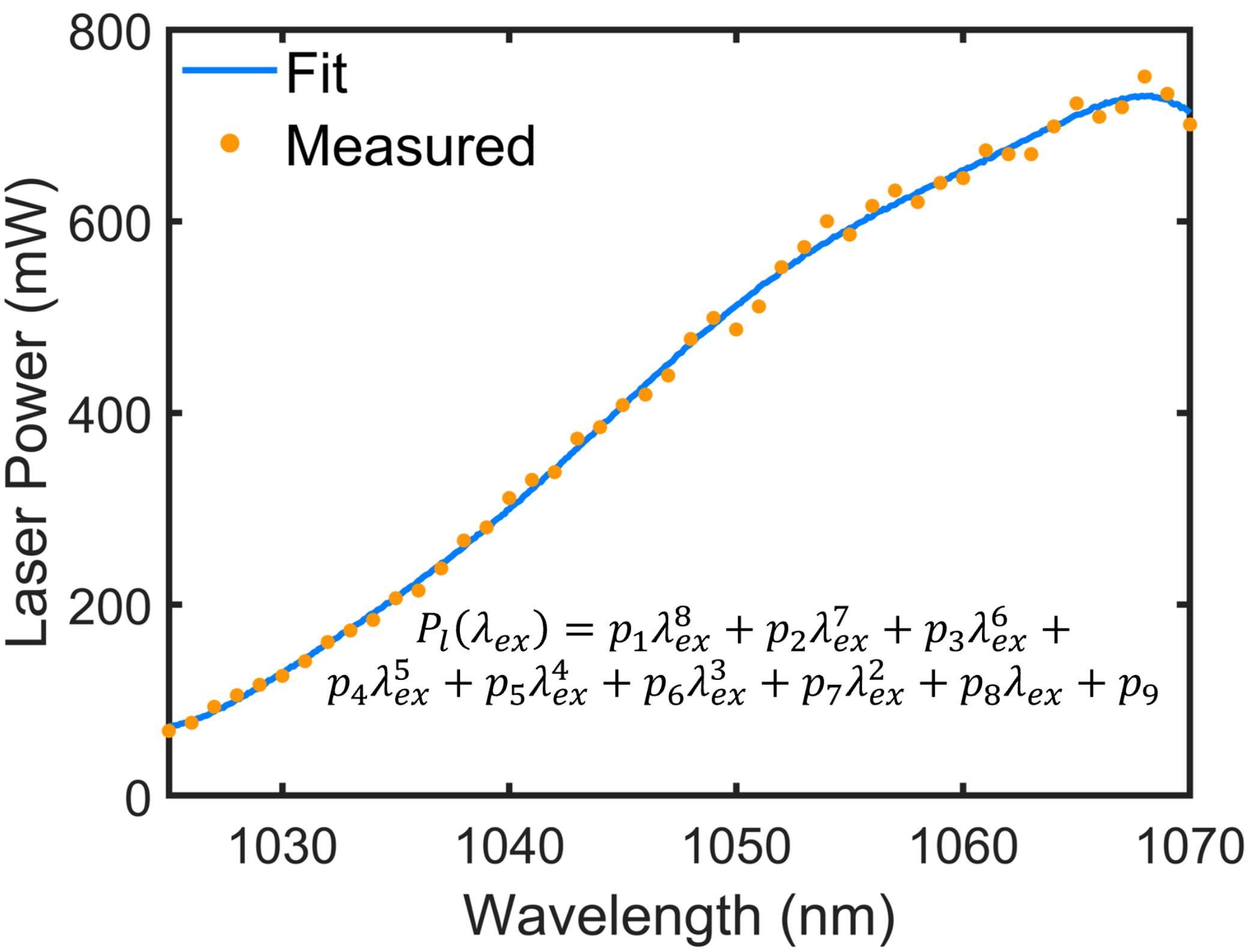

**Fig. S2. Fitting a polynomial equation to the power spectrum of the continuous wave tunable laser that drives the photothermal heating for numerical modelling.** $\lambda_{ex}$ is in units of nanometers and $P_l(\lambda_{ex})$ is in units of milliwatts. The power incident on the sample ($P_{ex}(\lambda_{ex})$) is $\approx 0.6 \times P_l(\lambda_{ex})$ due the beam splitter and other optical path losses. The fit parameters are given by:

- $p_1 = 3.8714\times10^{-11}$
- $p_2 =- 4.3487\times10^{-7}$
- $p_3 = 0.0020$
- $p_4 =- 5.0289$
- $p_5 = 7.6848\times10^{3}$
- $p_6 =- 7.3586\times10^{6}$
- $p_7 = 4.3335\times10^{9}$
- $p_8 =- 1.4399\times10^{12}$
- $p_9 = 2.0719\times10^{14}$

As can be seen, the laser power is nonuniform across the entire illumination range. Hence a normalized representation ($P_{norm} = \frac{P_{ex}(\lambda_{ex})}{P_{ex,max}(\lambda_{ex})}$) is used throughout the manuscript for representation. Following efforts have been made to counter the spectral variation of absolute illumination power ($P_{ex}(\lambda_{ex})$).

First, all models presented in Fig. 1 that are used to build the framework based on which all experimental data is interpreted, accounts for this power variation, to yield predictions that resemble experimental reality. The modeling in Fig. 1f generates phase portraits for different detunings at a constant normalized power. Their trends are in agreement with saddle node dynamics based theoretical predictions indicating that the detuning is the dominant contributor, thereby justifying the usage of normalized power representation for our analysis.

Second, our experiments in Fig. 2a, compare different excitation protocols at varying power levels. The comparisons are made between the behavior of different excitation protocols at the same detuning ($\Delta\lambda_0$) across the entire spectrum at different power levels ($P_{norm} = 1$). So, for each value of $P_{norm}$ the same spectral variation in absolute excitation power $P_{ex}(\lambda_{ex})$ will be in play. For example the peak temperature differential of 88°C between the forward spectral scan and static single wavelength scan is evaluated at the same detuning ($\Delta\lambda_0$) and the same normalized power ($P_{norm} = 1$) and therefore at the same absolute power $P_{ex}(\lambda_{ex})$. The same applies to the peak temperature differential of 72°C for $P_{norm} = 0.6$. This rationale also applies to Fig. 4a, and 6a as well as most of Fig. 2e and 6b (with the exception of a single cross wavelength comparison discussed in point 5). In fact, the bifurcation of the forward spectral scan and the single wavelength response into the low temperature branch as the laser wavelength redshifts, in line with theoretical predictions, despite that being accompanied by an increase in laser power, is a clear evidence that the spectral rate of change of laser power is not high enough to overshadow the dominant and intended detuning dependent effects. Most importantly, the spectral dependence of power cannot discount the claim that despite being illuminated by the same power at every detuning, the static single wavelength response bifurcates to the low temperature branch at a certain detuning, while the forward spectral scan continues to access high temperatures through bandgap penetration.

Third, Fig. 2b simply captures the spectral trend at different normalized power ($P_{norm}$) levels and therefore at different absolute power levels ($P_{ex}(\lambda_{ex})$). The comparison between the forward scan and static single wavelength response at $P_{norm} = 1.5$ compares their relative behavior at each detuning ($\Delta\lambda_0$) across the entire spectrum and the absolute power ($P_{ex}(\lambda_{ex})$) for each detuning is the same for both protocols. Similarly, Fig. 2c develops the mechanistic understanding of forward spectral scans through its laser detuning dependent jump across different static single wavelength excitation phase portraits. Both the forward spectral scan and static single wavelength response are generated for a range of detunings ($\Delta\lambda_0$) at the same normalized power level $P_{norm} = 1.5$ and therefore at the same absolute power ($P_{ex}(\lambda_{ex})$). In fact, the excellent overlay of the modeled trajectories for different excitation protocols indicates the detuning dependence dominates over the change in absolute power $P_{ex}(\lambda_{ex})$ across the detuning range.

Fourth, in Fig. 3, the ghost dynamics are captured over a very narrow spectral range for the spectral dependence of $P_{ex}(\lambda_{ex})$ to be consequential. Hence the -1/2 scaling law is faithfully achieved in Fig. 3b, 4b, 4c and 5c depict an increase in thermal bias required to drive interband transitions which is consistent with the theoretical predictions. The decrease in delay time with thermal bias at the same detuning does not depend on spectrally varying power levels. Similarly Fig. 5a and 5b involve analysis at a single illumination wavelength.

Fifth, the only cross wavelength quantitative comparisons that we make are the 13°C and 30°C difference in peak temperatures between the forward spectral scan and the static single wavelength response in Fig. 2e and 6b. Models using uniform excitation power ($P_{ex}$) in SI Note-12 for the bare silicon metasurface and in SI Note-17 for the gold integrated hybrid metasurface show that the central claim of increased temperature using spectral drag holds while the actual magnitude may differ.

## SI Note-7: Q-factor of metasurfaces derived by fitting asymmetric Fano function to ellipsometric transmittance data

The following function was used to fit the transmittance data from the ellipsometer[S6]:

$$T\left(\omega_{ex}\right) = a + b\frac{\left(q+\frac{\omega_{ex}-\omega_0}{\frac{\Gamma}{2}}\right)^2}{1+\left(\frac{\omega_{ex}-\omega_0}{\frac{\Gamma}{2}}\right)^2} \tag{S32}$$

Here, $a$ is the background transmittance, $b$ is the resonator enhancement, $q$ is the asymmetry parameter, $\omega_0$ is the resonance frequency which is given by $\omega_0 = \frac{2\pi c}{\lambda_0}$. $\Gamma = \frac{\omega_0}{Q}$ is the full width at half maxima with $Q$ being the Q-factor of the metasurface. $Q$ is extracted from the fits and used in experimental interpretation as well as numerical modelling.

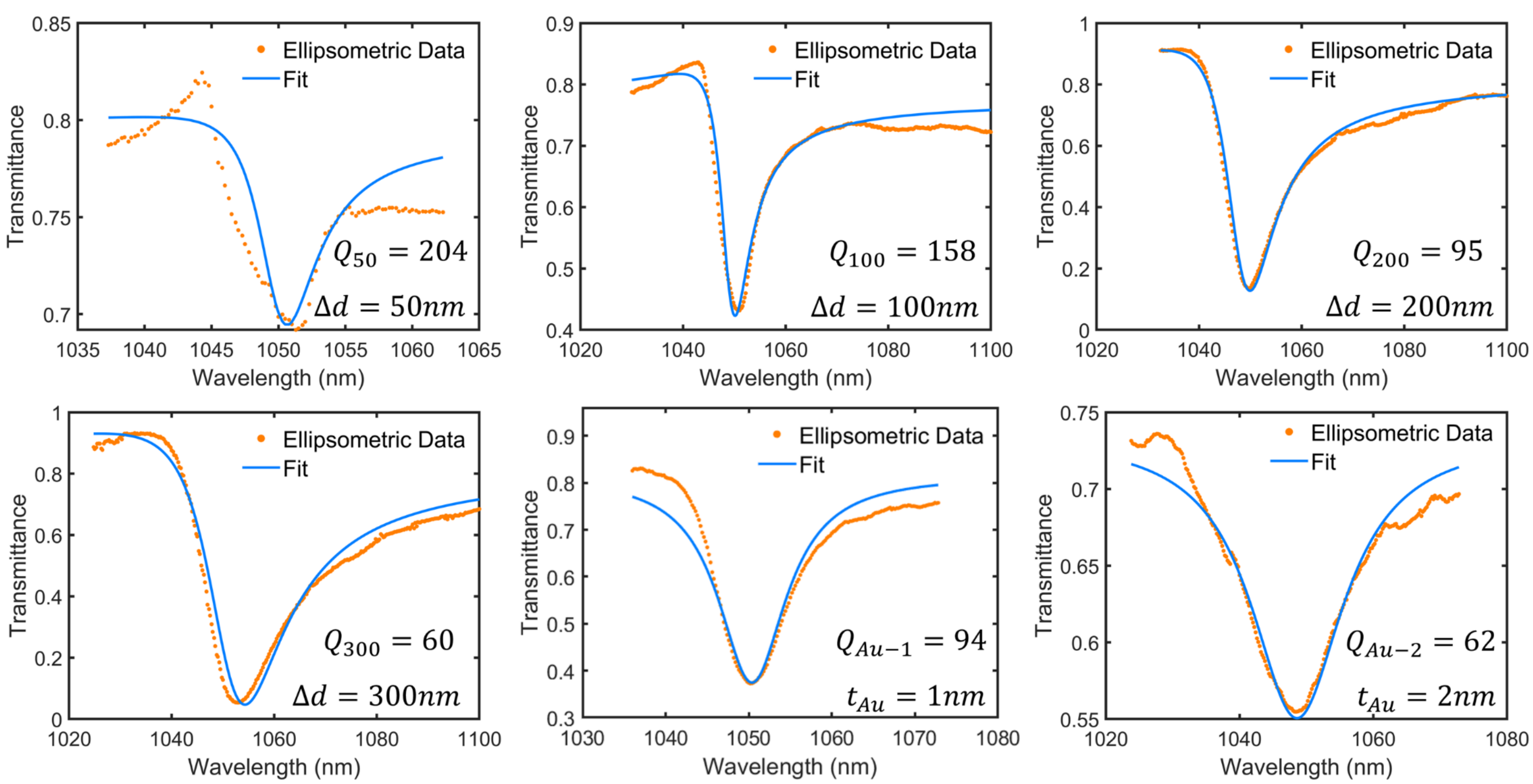

**Fig. S3. Q-factor extraction from ellipsometric transmittance data by fitting asymmetric fano function.**

**SI Note-8: Simulation of the mode field distribution of the bare amorphous silicon metasurface with $\Delta d = 100\ nm$**

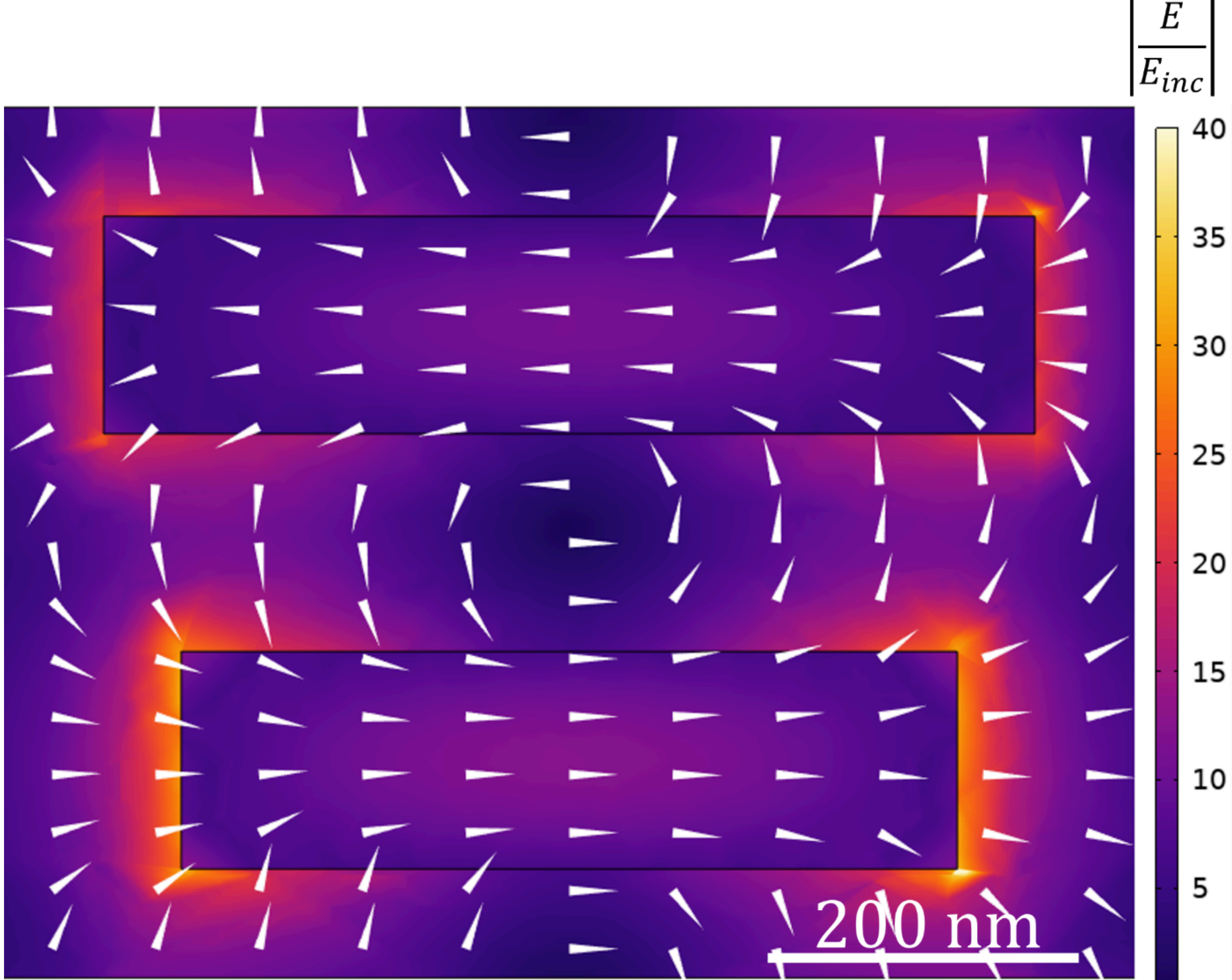

**Fig. S4. Resonant mode profile of biperiodic amorphous silicon metasurface on fused silica.** The simulated structures had a mean length of $d = 550\ nm$ with a reference perturbation of $\Delta d = 100\ nm$. Both the block width $t_{Si}$ and the inter block gap ($t_{gap}$) are $140\ nm$. The block heights ($h_{Si}$) were chosen to be $112\ nm$ to obtain the above resonant mode at $\approx\ 1050\ nm$.

**SI Note-9: Schematic of experimental setup**

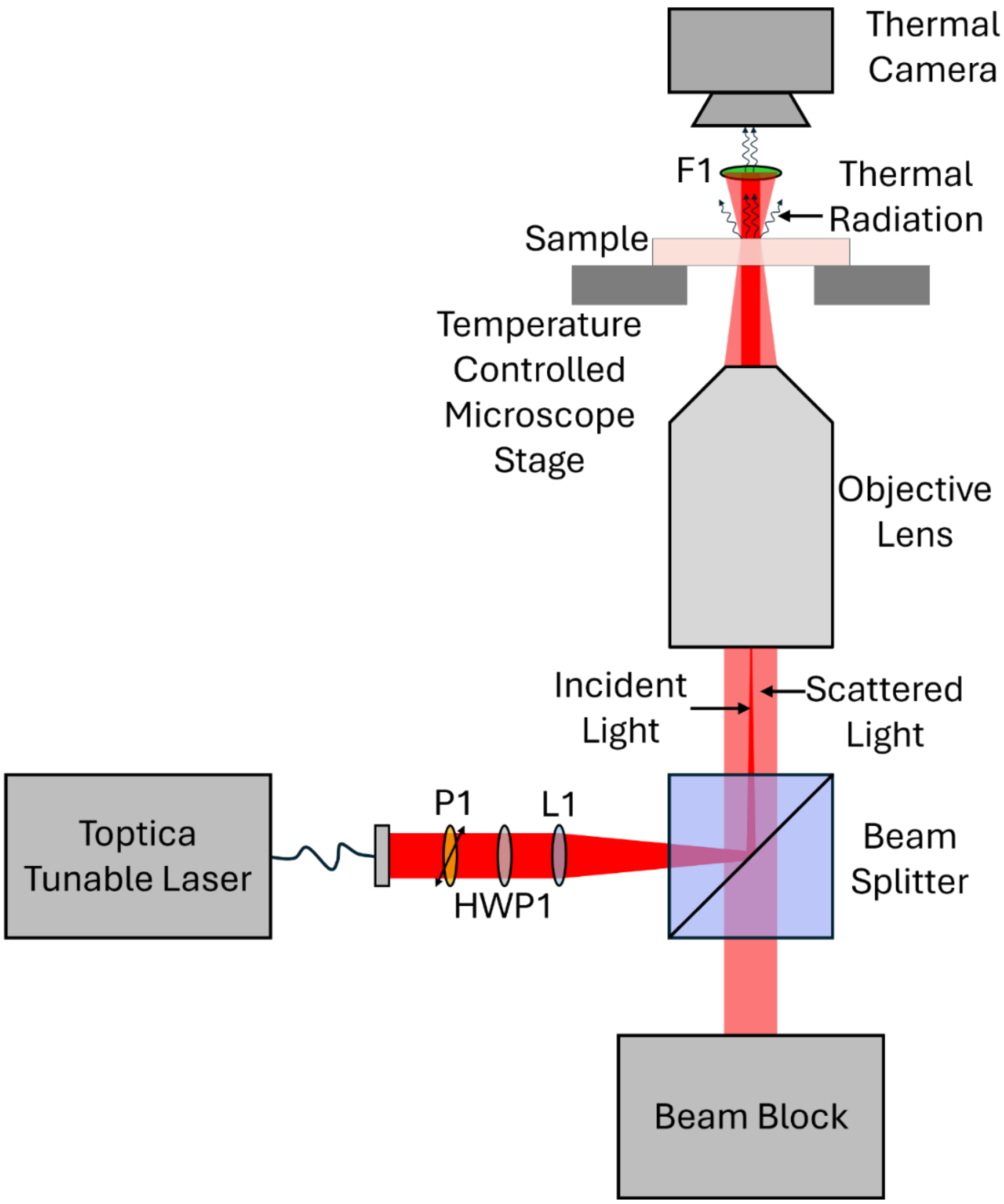

**Fig. S5. Complete experimental setup for photothermal measurements.** A Toptica tunable laser along with a booster is used as the source. The collimated output out of the fiber passes through a linear polarizer (P1) following which a half wave plate (HWP1) is used to rotate its polarization to line up the direction of the incident electric field along the length of the silicon nanobars. A 125 mm biconvex lens (L1) focusses the incident light on the back focal plane of the 20x objective (OB), with 0.45 numerical aperture (N.A.) and a working distance between $8.18\,mm$ - $8.93\,mm$, after passing through a beam splitter (BS). The objective (OB) then illuminates the metasurface mounted on a x-y translation stage with a near parallel beam of light. The transmitted beam is filtered out by a long pass filter (F1) with a pass band from $7.6\ \mu m$ - $14.6\ \mu m$. Only the thermal radiation within this wavelength range reaches the lens of the infrared/thermal camera which then focuses it on the detector to provide the studied thermal information. The reflected beam is blocked off by a beam block.

**SI Note-10: Experiment and Model Comparison of Single Wavelength, Forward and Backward Scan Trajectories, Temperatures and Bifurcation Shifts**

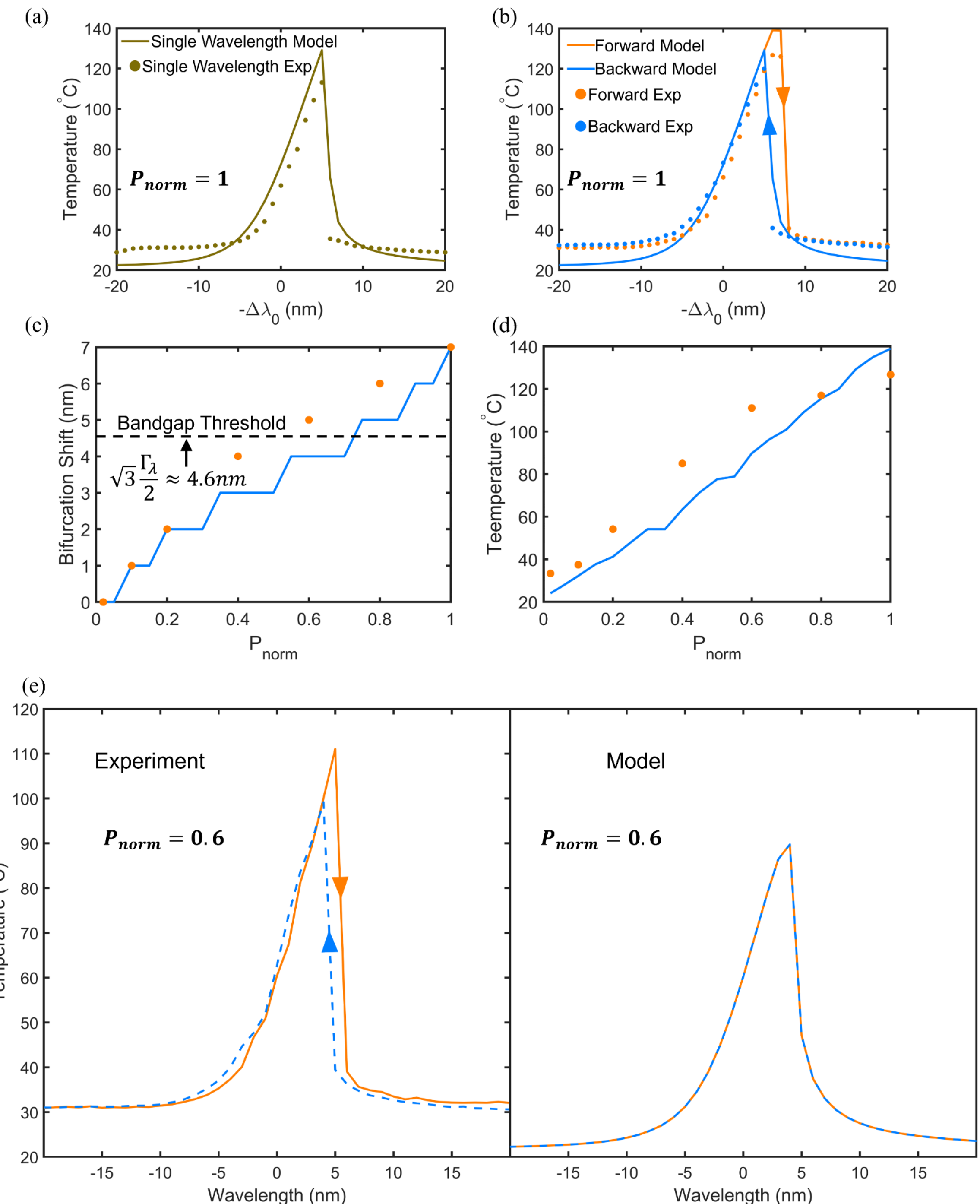

**Fig. S6. Quantitative comparison between thermal balance model prediction and experiments. (a)** Comparison between thermal balance model and single wavelength excitation experiments at $P_{norm} = 1$, **(b)** and forward & backward spectral scans at $P_{norm} = 1$. **(c)** Spectral shift in the high temperature branch to low temperature branch bifurcation point with $P_{norm}$ for forward scans, expressed in terms of the initial detuning ( $-\lambda_0$). The dashed black line represents the estimated detuning threshold for the emergence of bandgap from the normalized thermal balance model (SI Note-4). For smaller $P_{norm}$ the photothermal response maintains a largely symmetric lineshape. So, the bifurcation shift would simply be the resonance shift. As $P_{norm}$ increases, the photothermal response assumes an increasingly asymmetric lineshape along with a larger resonance shift. $P_{norm} = 0.6$ yields bifurcation shifts close to the band formation threshold. As the model estimated and experimentally obtained bifurcation point at $P_{norm} = 0.6$ lie on either side of the threshold line, they yield different hysteresis widths. **(d)** Maximum temperatures of the forward spectral scan with varying $P_{norm}$. **(e)** Experimental forward and backward spectral scan data (left panel) at $P_{norm} = 0.6$ yields a hysteresis width of $1\ nm$ as it is above the threshold line for bandgap emergence while the model data exhibits no hysteresis as it is below the threshold line for bandgap emergence. The discrepancy arises due to multiple estimated model parameters and variation in laser power.

**SI Note-11: Simulations confirming that the spectral drag generates maximum heating compared to switched scan and power drag**

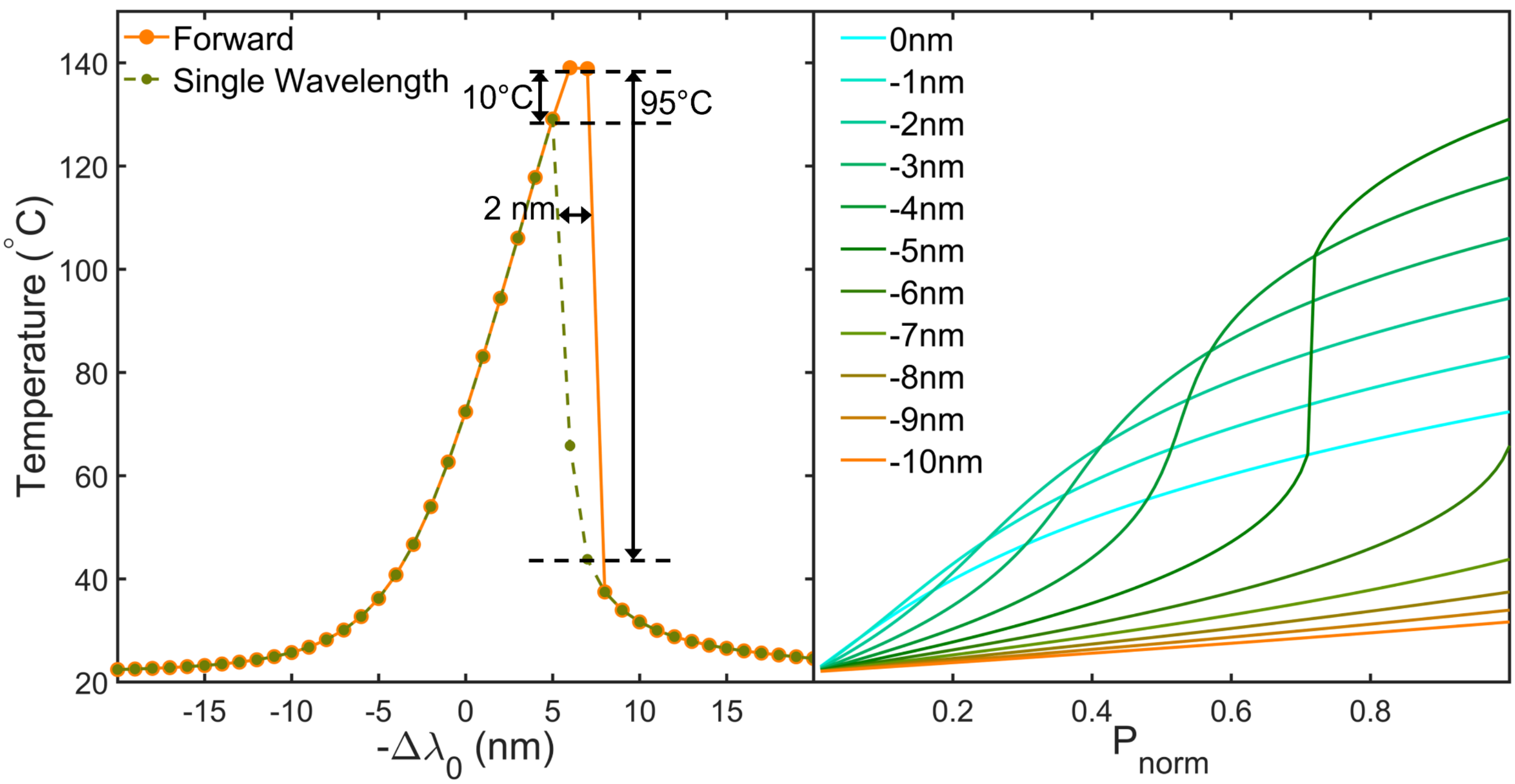

**Fig. S7. Numerical simulations of the photothermal response of the metasurface with spectral scan and power scan.** It supports experimental data in Fig. 2e that the maximum heating is obtained from spectral drag. As can be seen in the left panel temperature due to continuous spectral scan is 10℃ higher compared to static single wavelength excitation at $P_{norm} = 1$ (experimental data in Fig. 2e shows a 13℃ difference). Even comparing with the continuously increasing power scan in the right panel, we see that spectral drag in forward spectral scan produces maximum heating. The peak temperature differential between forward scan and static single wavelength excitation shown here is 95℃ (experimental data in Fig. 2a shows a 88℃ differential). Modelling in SI Note-12 shows that the same qualitative behavior is achieved even when excitation power is uniform across the excitation spectrum.

**SI Note-12: Nonlinear photothermal spectra of forward spectral scan and static single wavelength excitation under idealized scenario of uniform illumination power across the spectrum using simulations.**

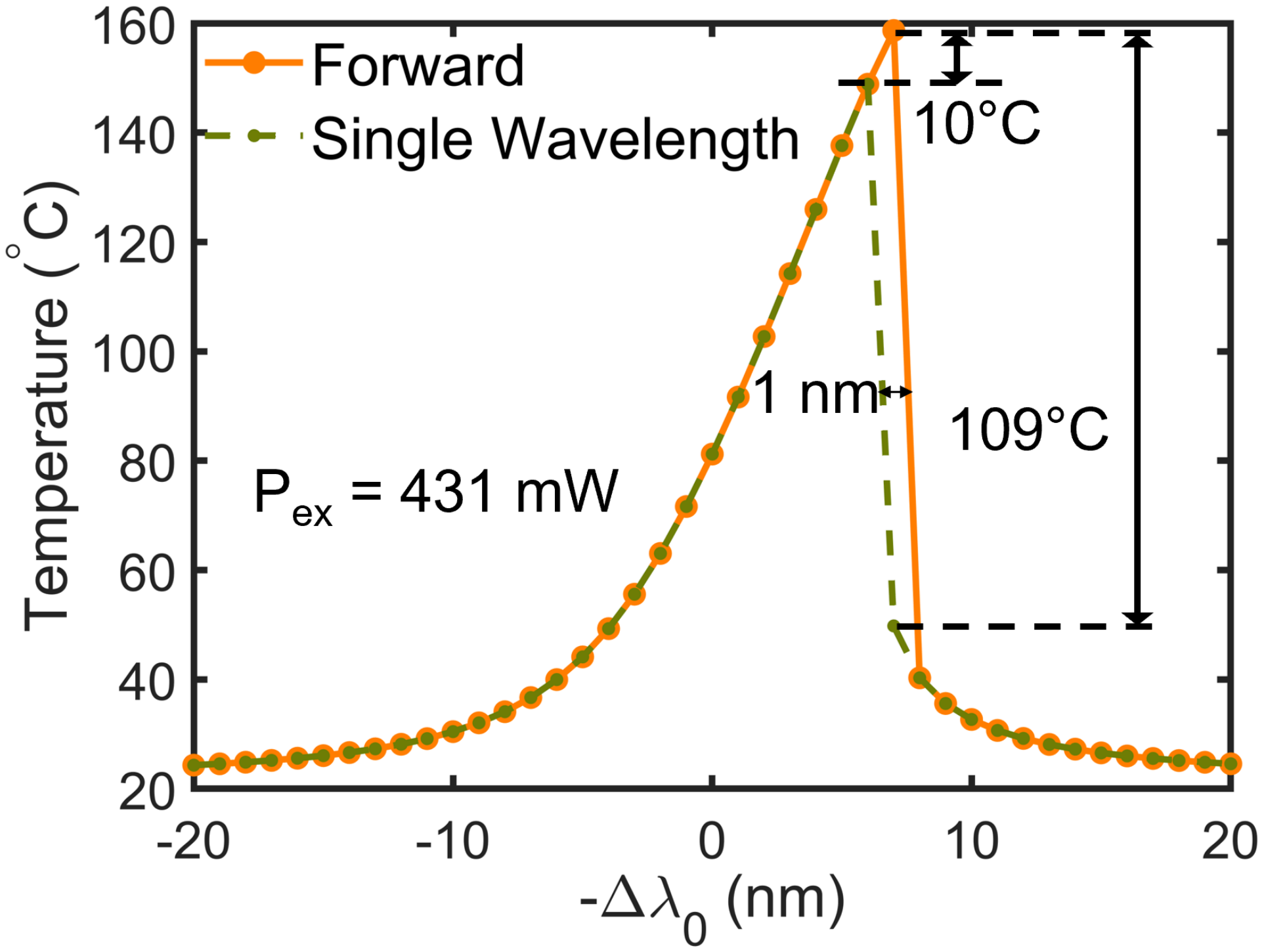

**Fig. S8. Uniform power simulations of the photothermal response of the metasurface under continuous forward scan and static single wavelength single excitation.** The spectrally uniform illumination power of $P_{ex} = 431\ mW$ is chosen to ensure that the forward spectral scan bifurcates from the high temperature branch to the low temperature branch at $\Delta\lambda_0 = -\ 8\ nm$. It is $13.6\%$ more than the maximum laser spectrum power of $379\ mW$ at $\Delta\lambda_0 = -\ 7\ nm$, the edge of the high temperature solution branch. While the peak heating in forward scan is 10°C higher compared to static single wavelength excitation (same as the laser spectrum model), the peak temperature differential between the forward spectral scan and static single wavelength response is 109°C (slightly higher than 95°C for the laser spectrum model) and the hysteresis width is $1\ nm$ (slightly lower than $2\ nm$ for the laser spectrum model). The results are qualitatively similar and exhibit the same spectral and power dependencies and same behavior. Refer to SI Note-17, for spectrally uniform illumination modelling of metasurfaces with $1\ nm$ film of gold.

**SI Note-13: Thermoelectric stage $T(t)$ vs $V$**

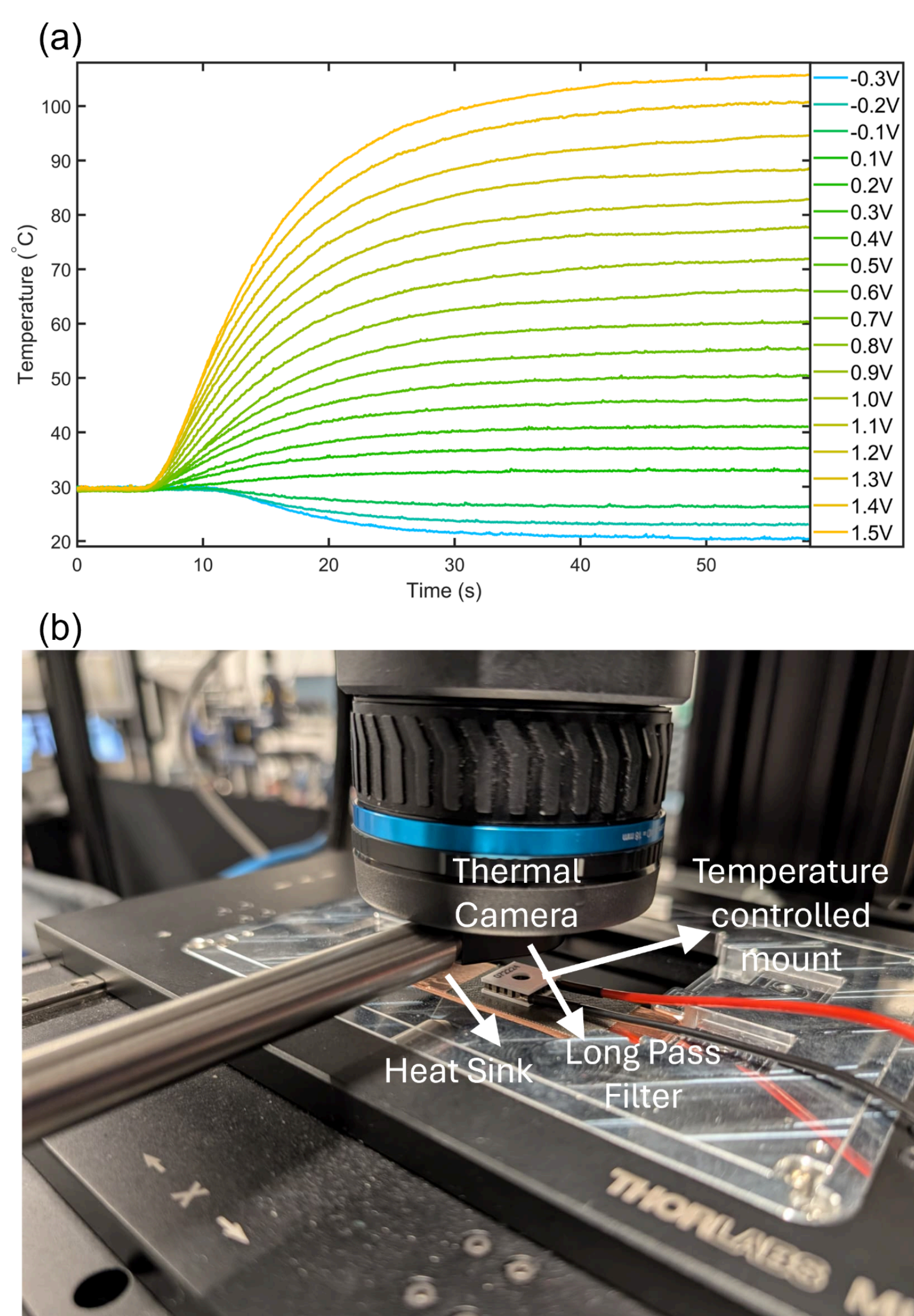

**Fig. S9. (a) Steady state temperature of the sample stage with varying voltage applied across the terminals of the thermo-optic sample mount. (b) Experimental setup with the temperature-controlled sample stage.**

## SI Note-14: Thermal measurement error bounds

***Thermal camera sampling error***

As seen from the experimental data, the camera frame rate suffices to measure the slow temperature evolution data. To estimate the temperature measurement error due to camera sampling rate associated with the fast interband transitions, which is crucial for accurate estimation of the band edge values we look at:

$$\delta T_{sampling} = \left|\frac{\partial T}{\partial t}\right|_{transition} \times \delta t_{sampling}$$

(S33)

From the experimental data we know $\delta t_{sampling} = 33\, ms$. Also, knowing that the temperature transition between two successive frames during the fast transition is $\approx 50\,°C$ (Fig. 3 & 4), we obtain $\left|\frac{\partial T}{\partial t}\right|_{transition} \approx \frac{50°C}{\delta t_{sampling}}$. So, for bare aSi metasurface $\delta T_{sampling,max} \approx 50\,°C$. This represents the worst case variability in band edge temperature estimates.

***Thermo-optic coefficient and resonance linewidth estimation error***

From the normalized analysis of the photothermal nonlinearity in SI Note-4 we know that the band edge temperature ($\Delta T_{BE}$) depends on the thermo-optic shift coefficient ($\alpha$) and the resonance linewidth ($\frac{\Gamma_\omega}{2}$). The error in $\Delta T_{BE}$ can be estimated as:

$$\delta \Delta T_{BE} = \left|\frac{\partial \Delta T_{BE}}{\partial \frac{\Gamma_\omega}{2}}\right| \delta \frac{\Gamma_\omega}{2} + \left|\frac{\partial \Delta T_{BE}}{\partial \alpha}\right| \delta\alpha$$

(S34)

Knowing the form of the normalized band edge temperature in Eq. S23 we can write:

$$\left|\frac{\partial \Delta T_{BE}}{\partial \frac{\Gamma_\omega}{2}}\right| = \frac{\frac{\Gamma_\omega}{2}}{|\alpha|\sqrt{\delta_0^2 - 3\left(\frac{\Gamma_\omega}{2}\right)^2}} \qquad \text{(S35)}$$

$$\left|\frac{\partial \Delta T_{BE}}{\partial \alpha}\right| = \frac{2\delta_0 + \sqrt{\delta_0^2 - 3\left(\frac{\Gamma_\omega}{2}\right)^2}}{3|\alpha|^2} \qquad \text{(S36)}$$

Here, $\delta_0 = \omega_0 - \omega_{ex} = 2\pi c\left(\frac{1}{\lambda_0} - \frac{1}{\lambda_{ex}}\right)$. So, for $\lambda_0 = 1050\, nm$ and $\lambda_{ex} = 1055\, nm$ ($\Delta\lambda_0 =- 5\, nm$), $\delta_0 = 8.5 \times 10^{12}\, Hz$; for $Q = 198$, $\frac{\Gamma_\omega}{2} = 4.53 \times 10^{12}\, Hz$ and $|\alpha| = 9.3 \times 10^{10}\, Hz/°C$ (SI Note-2). Using these values in Eq. S35 and S36 we obtain: $\left|\frac{\partial \Delta T_{BE}}{\partial \frac{\Gamma_\omega}{2}}\right| = 1.49 \times 10^{-11}\, °C/Hz$ and $\left|\frac{\partial \Delta T_{BE}}{\partial \alpha}\right| = 7.81 \times 10^{-10} °C^2/Hz$. For an experimentally measured $Q = 158$, $\delta\frac{\Gamma_\omega}{2} = 1.15 \times 10^{12}\, Hz$. Again we have assumed $\frac{\partial n_{Si}}{\partial T} = 3.2 \times 10^{-4}$. Reported thermo-optic coefficient values for amorphous silicon span between $2 \times 10^{-4} - 4 \times 10^{-4}/K$. While the lower end is the typical value around $\approx 1550\, nm$ excitation, the upper end is the typical value for $< 1000\, nm$ illumination. So, we conservatively assume a maximum variation of $1.2 \times 10^{-4}$, i.e. $\delta\alpha = 1.2 \times 10^{-4} \times \frac{\partial \omega_{res}}{\partial n_{Si}} = 3.49 \times 10^{10}\, Hz/°C$. Using these values in Eq. S34 we obtain $\left|\frac{\partial \Delta T_{BE}}{\partial \frac{\Gamma_{\omega}}{2}}\right| \delta\frac{\Gamma_{\omega}}{2} = 17.06°\text{C}$ and $\left|\frac{\partial \Delta T_{BE}}{\partial \alpha}\right| \delta\alpha = 27.25°\text{C}$. As $\delta\alpha$ is a conservative estimate starting towards the upper bound, the latter error can be much smaller.

The elevated substrate temperature under external bias adds a systematic offset to the absolute band-edge temperatures not accounted for in the model, while camera sampling and parameter estimation contribute additional uncertainty. Together they account for the discrepancies reported in the main text.

**SI Note-15: Deriving amplification and time delay scaling law using ghost dynamics with additional thermal bias**

The ghost dynamics near the band-edge are given by:

$$\frac{\partial \Delta T}{\partial t} = \mu + \Delta T^2 \quad \text{(S37)}$$

Here, $\mu = \Delta\lambda_0 - \Delta\lambda_{BE}$. When an external thermal bias is introduced the thermo-optic shift in the resonance leads to a modified effective $\mu$ ($\mu_{eff}$),

$$\mu_{eff} = \Delta\lambda_0 - \Delta\lambda_{BE} + \psi\Delta T_{bias} \quad \text{(S38)}$$

From this we know that,

$$\mu_{eff} \propto \Delta T_{bias} \quad \text{(S39)}$$

Now looking at the process of thermal amplification, $\Delta T_{amp}$ is the ratio of temperature rise with and without the thermal bias ($\Delta T_{amp} = \frac{\Delta T_{high}}{\Delta T_{low}}$). Therefore, the normalized amplification gain ($G_{norm}$), where $\Delta T_{amp}$ is normalized by the bias temperature rise ($\Delta T_{bias}$) can be written as:

$$G_{norm} = \frac{\Delta T_{amp}}{\Delta T_{bias}} \quad \text{(S40)}$$

This makes $G_{norm} \propto \frac{1}{\Delta T_{bias}}$. Using Eq. S39 then we can write:

$$G_{norm} \propto \frac{1}{\mu_{eff}} \quad \text{(S41)}$$

From the main text it is known that $t_{Delay} \propto \frac{1}{\sqrt{\mu_{eff}}}$. Using this relation in Eq. S41 we can write:

$$G_{norm} \propto t_{Delay}^2 \quad \text{(S42)}$$

From Eq. S42 we can safely write:

$$t_{Delay} \propto \sqrt{G_{norm}} \quad \text{(S43)}$$

However, this needs the underlying assumption that $t_{Delay} \propto \frac{1}{\sqrt{\mu_{eff}}}$. This relation is only valid for a constant $\mu_{eff}$ (as per our derivation in the main text). This requires $\Delta T_{bias}$ to be time invariant. However, as shown in SI Note-13, $\Delta T_{bias}$ requires time comparable to $t_{Delay}$ to reach a steady state. Therefore, the relationship in Eq. S43 is a good approximate relation.

**SI Note-16: Variation in absorption efficiency and Q-factor with change in absorption and scattering**

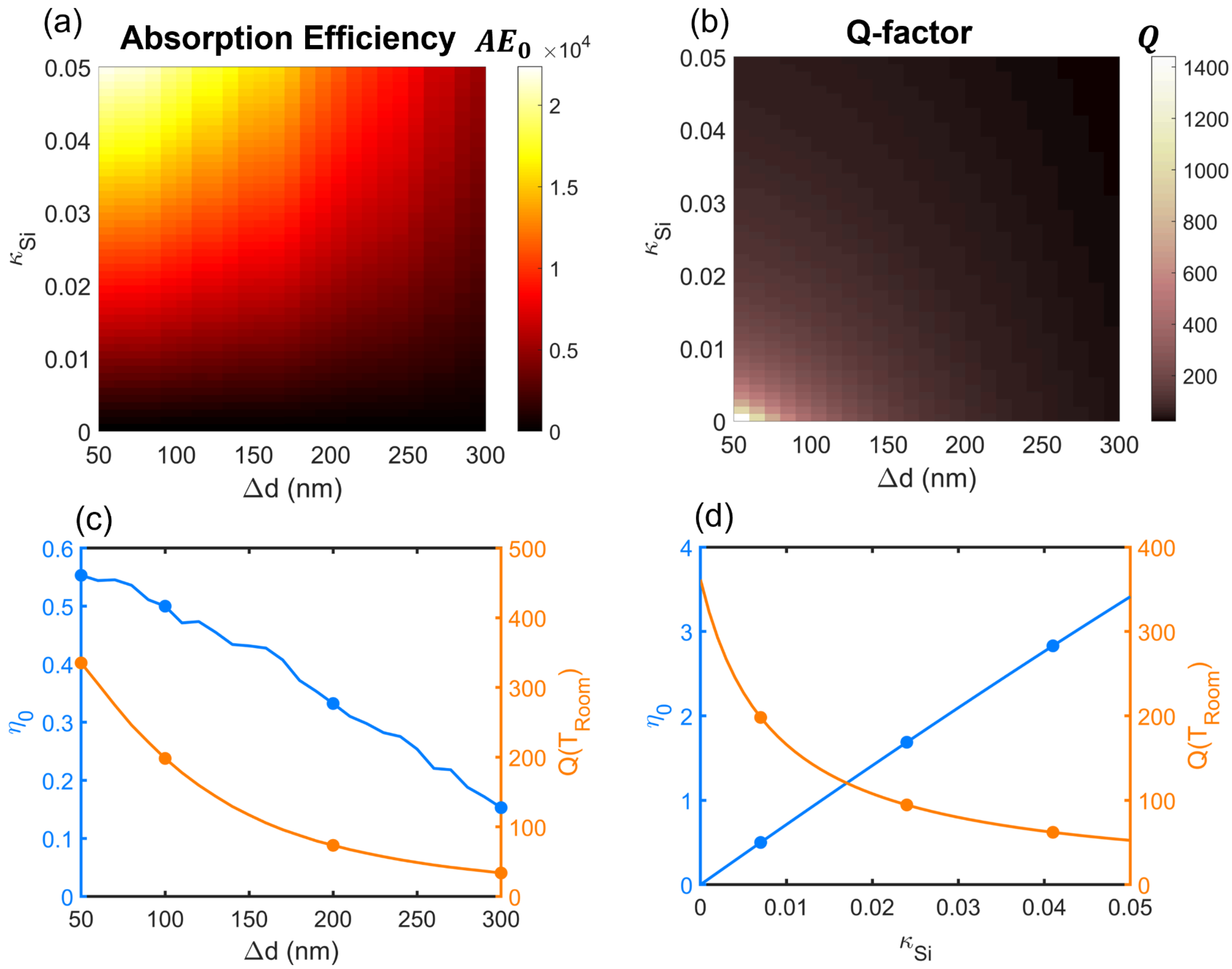

**Fig. S10. Variation in metasurface absorption and Q-factor with change in absorption and scattering.** (a) Absorption efficiency increases due to increased photon confinement within the cavity with decreased scattering (decrease in $\Delta d$) and increase in absorption. (b) Q-factor increases with decrease in photon scattering and increase in confinement while it decreases with increase in material absorption. (c) Combining (a) and (b) we see that the normalized absorption fraction ($\eta_0$) of the metasurface decreases with decrease in $Q$ due to increased scattering. This holds for when operating in the scattering dominated loss side of critical coupling. (d) On the other hand, he normalized absorption fraction ($\eta_0$) of the metasurface increases with decrease in $Q$ due to increased absorption. It explains the increased heating with the incorporation of gold but decreased heating with increase in scattering (increase in $\Delta d$).

**SI Note-17: Thermo-optic response of amorphous silicon–gold hybrid systems.**

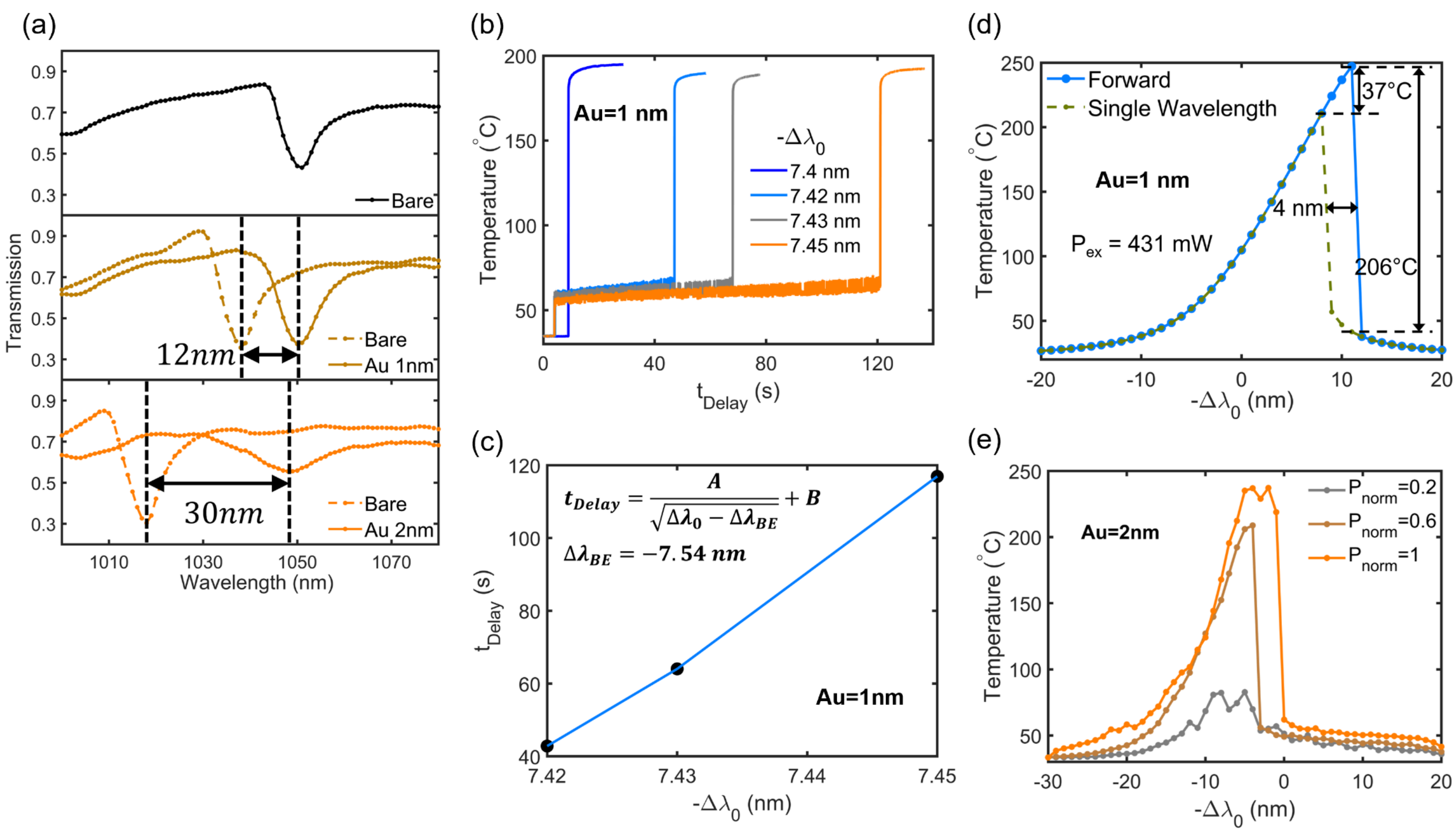

**Fig. S11. Spectral response of amorphous silicon-gold hybrid metasurfaces.** (a) As the thickness of gold film increases the metasurface resonance exhibits a larger redshift. Initial resonances of bare silicon metasurfaces are chosen to achieve a resonance of $\approx 1050\,nm$, after the deposition of gold film. (b) Ghost dynamics in aSi metasurfaces with $1\,nm$ thin film of gold for $P_{norm} = 1$. (c) The -1/2 power law scaling of $t_{Delay}$ with distance from the band edge in line with the bare aSi metasurface. It helps determine the band edge without knowledge of all the relevant material parameters for the hybrid system. (d) The simulated spectral response of the metasurface with $1\,nm$ thin film of gold for spectrally uniform illumination power of $431\,mW$ which is 6% more than the $404\,mW$ laser spectrum power at $\Delta\lambda_0 = -11\,nm$ (the experimentally observed bifurcation point for the metasurface in Fig. 6b of main text). The model exhibits the same bifurcation point as well as the same width ($4\,nm$) of spectral drag of the forward spectral scan compared to the static single wavelength response. The peak temperature differential of 206°C between the forward scan and static response as well as the difference in peak temperatures of 37°C are slightly higher than the experimentally reported values (169°C and 30°C respectively) due to the higher power. Nevertheless, the results are qualitatively similar yielding similar responses. (e) For a gold film of $2\,nm$, the peak heating and subsequent bifurcation to the low temperature state is achieved at $0\,nm$ detuning for an excitation power of $P_{norm} = 1$. At lower powers, the peak is achieved at blue detunings. This indicates the significant contribution of the temperature dependent optical properties of gold to the photothermal response of the hybrid metasurface.

**SI Note-18: Thermal camera characterization for spatial resolution**

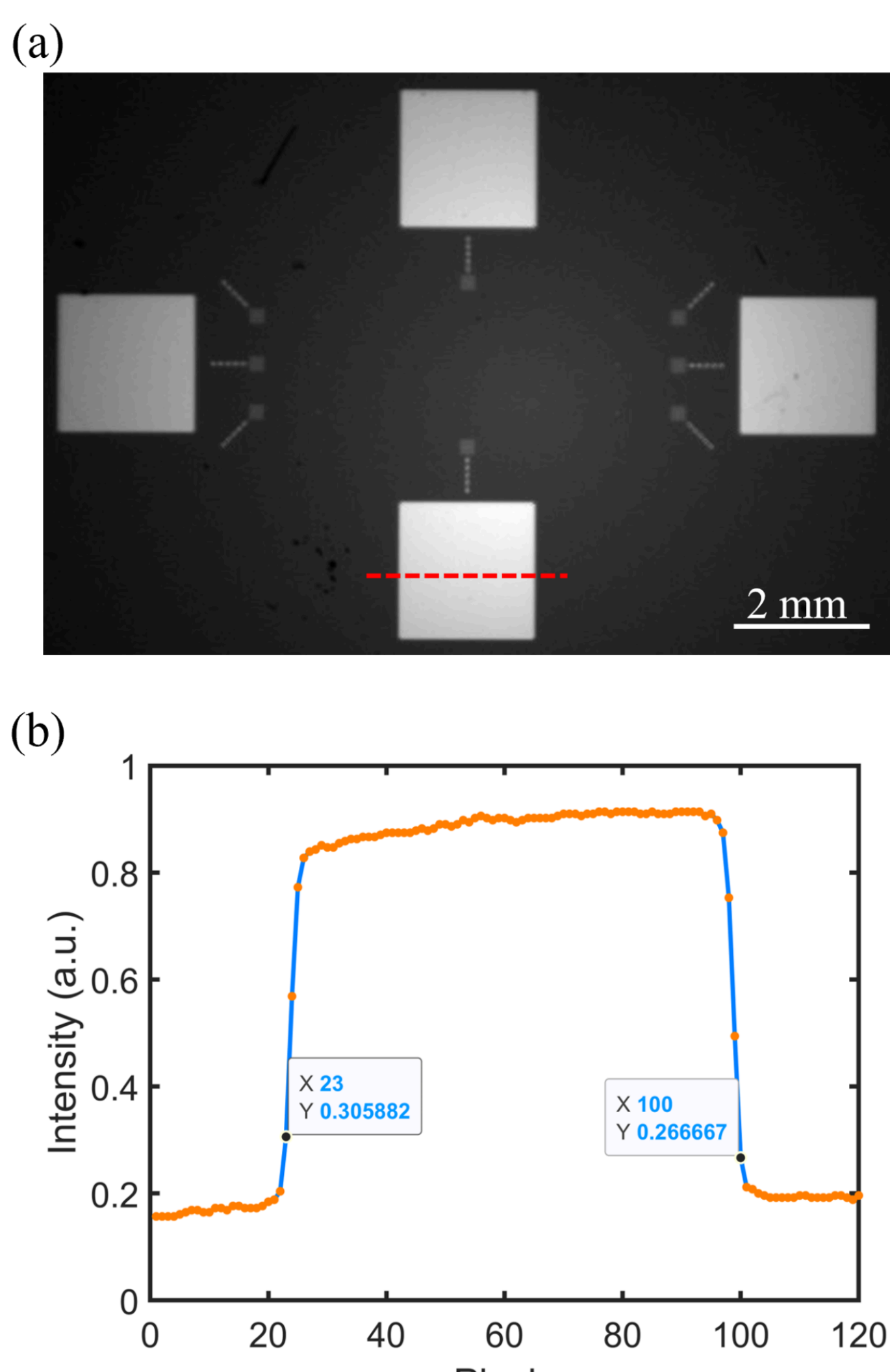

**Fig. S12. Spatial resolution of thermal imaging. (a)** Thermal camera calibration using a characterization sample consisting of gold structures on fused silica. The large gold pads are $2\,mm$. **(b)** The different emissivities of gold and fused silica are seen as different colors on the camera. Measurements along the red line shows that the gold pad spans 78 pixels which corresponds to a spatial resolution of $\frac{2000\ \mu m}{78} = 25.64\ \mu m$ which is 4 times more than the beam diameter of $\approx\ 100\ \mu m$.

**SI Note-19: Dependence of the metasurface resonance and Q-factor on the real and imaginary part of the refractive index**

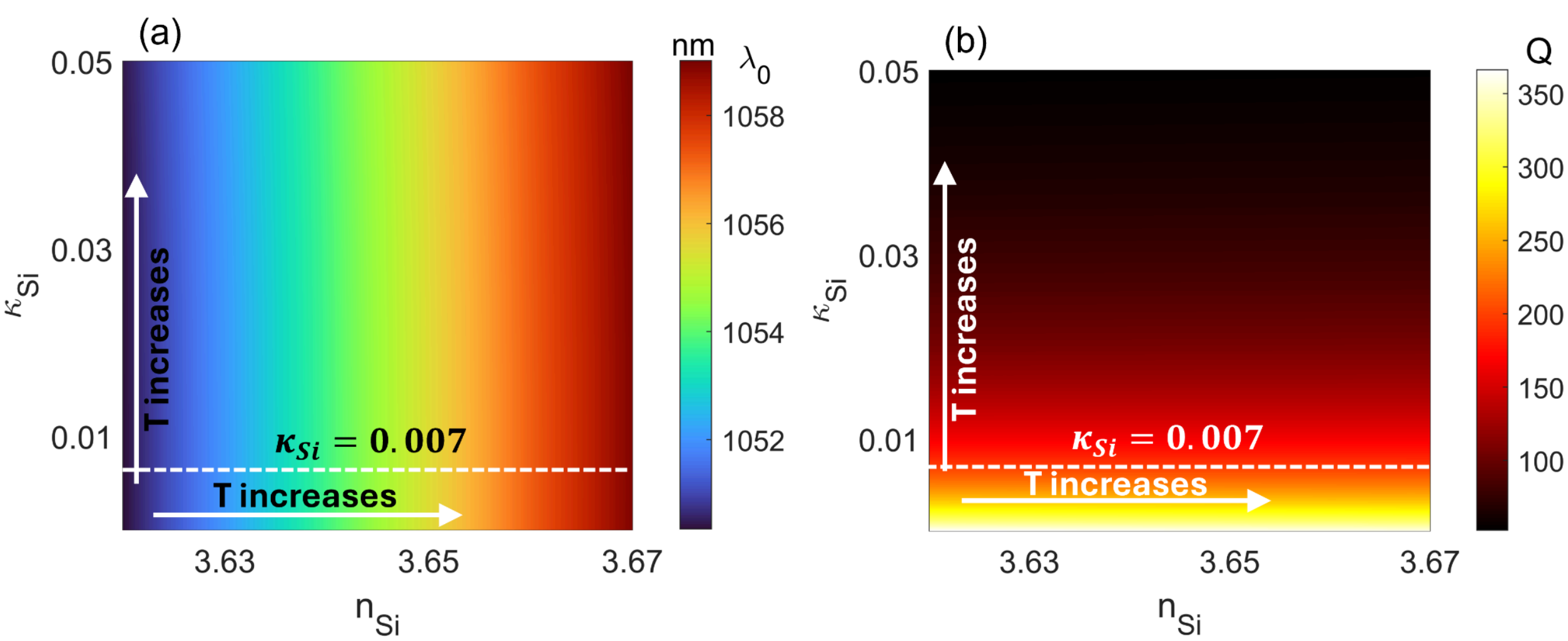

**Fig. S13. Dependence of the metasurface resonance lineshape on the real and imaginary part of the refractive index.** The real part predominantly impacts the resonance wavelength ($\lambda_0$), while the imaginary part primarily determines its Q-factor ($Q$). The positive thermo-optic coefficient of both the real and imaginary parts drive the observed photothermal nonlinearity of the amorphous silicon metasurface. A nominal $\kappa_{Si} = 0.007$ has been assumed for the amorphous silicon for all thermal balance modelling (under the lumped approximation) for close match with experimentally observed temperatures and bistable transition points.

## SI Note-20: Extracting $\frac{\partial\lambda_{res}}{\partial n_{Si}}$ and $\frac{\partial\omega_{res}}{\partial n_{Si}}$ from COMSOL Simulations

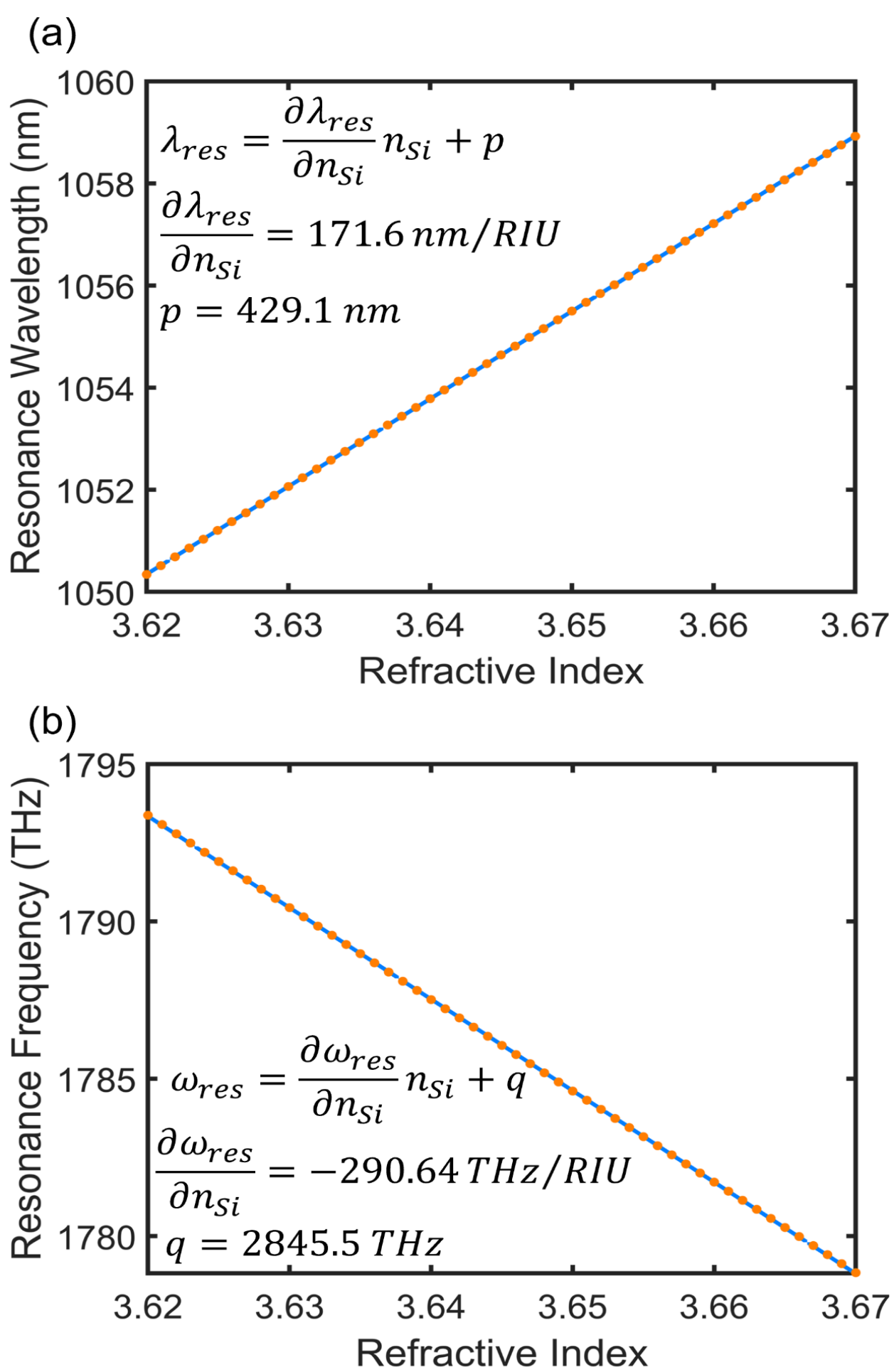

**Fig. S14. Extracting $\frac{\partial\lambda_{res}}{\partial n_{Si}}$ and $\frac{\partial\omega_{res}}{\partial n_{Si}}$ from COMSOL simulations of the change in $\lambda_{res}$ with change in $n_{Si}$.** Using linear curve fits to the numerically obtained function for $\lambda_{res}(n_{Si})$ we extract $\frac{\partial\lambda_{res}}{\partial n_{Si}}$ and $\frac{\partial\omega_{res}}{\partial n_{Si}}$. It is to be noted that $\lambda_{res} = \lambda_0$ and $\omega_{res} = \omega_0$ at room temperature. As we can $\frac{\partial\omega_{res}}{\partial n_{Si}} < 0$. This makes $\alpha < 0$.

## SI Note-21: Extracting $Q\left(\kappa_{Si}\right)$ from COMSOL Simulations

$$\frac{\partial Q}{\partial T} = \frac{\partial Q}{\partial n_{Si}}\frac{\partial n_{Si}}{\partial T} + \frac{\partial Q}{\partial \kappa_{Si}}\frac{\partial \kappa_{Si}}{\partial T} \tag{S44}$$

From SI Note-19 it is known that

$$\frac{\partial Q}{\partial \kappa_{Si}} \gg \frac{\partial Q}{\partial n_{Si}} \tag{S45}$$

Hence, change in $Q$ can be simplified as

$$\frac{\partial Q}{\partial T} \cong \frac{\partial Q}{\partial \kappa_{Si}}\frac{\partial \kappa_{Si}}{\partial T} \tag{S46}$$

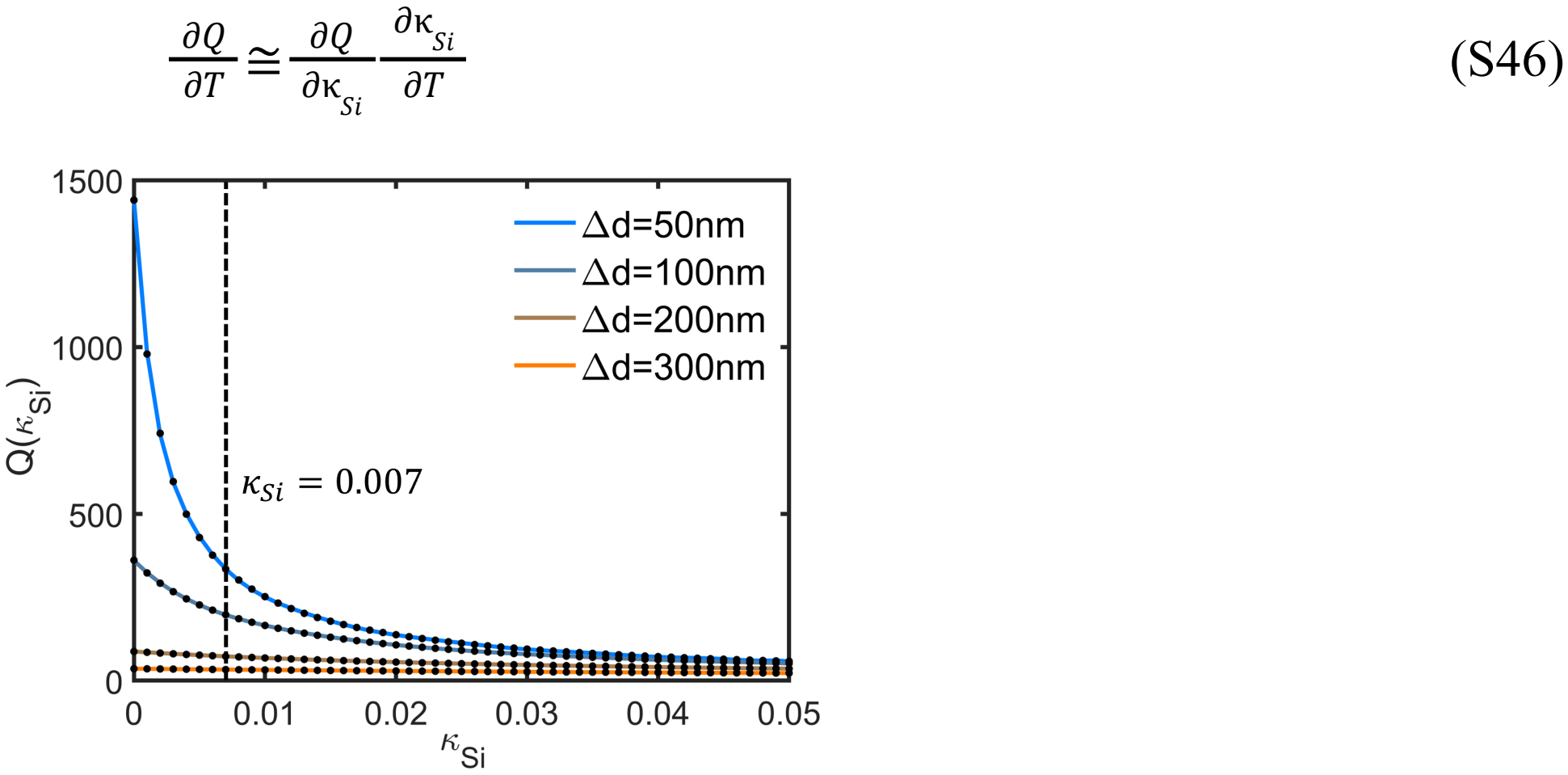

| | $\Delta d = 50nm$ | $\Delta d = 100nm$ | $\Delta d = 200nm$ | $\Delta d = 300nm$ |
|---|---|---|---|---|
| $p_{13}$ | $-6.4938 \times 10^{24}$ | | | |
| $p_{12}$ | $2.2368 \times 10^{24}$ | | | |
| $p_{11}$ | $-3.4398 \times 10^{23}$ | | | |
| $p_{10}$ | $3.1163 \times 10^{22}$ | | | |
| $p_{9}$ | $-1.8481 \times 10^{21}$ | | | |
| $p_{8}$ | $7.5425 \times 10^{19}$ | | | |
| $p_{7}$ | $-2.1683 \times 10^{18}$ | $-9.9372 \times 10^{12}$ | | |
| $p_{6}$ | $4.4222 \times 10^{16}$ | $2.0150 \times 10^{12}$ | | |
| $p_{5}$ | $-6.3679 \times 10^{14}$ | $-1.6910 \times 10^{11}$ | | |
| $p_{4}$ | $6.3816 \times 10^{12}$ | $7.6493 \times 10^{9}$ | $5.0347 \times 10^{6}$ | |
| $p_{3}$ | $-4.3579 \times 10^{10}$ | $-2.0526 \times 10^{8}$ | $-7.9333 \times 10^{5}$ | $-2.1228 \times 10^{4}$ |
| $p_{2}$ | $1.9907 \times 10^{8}$ | $3.4589 \times 10^{6}$ | $5.4492 \times 10^{4}$ | $3.9477 \times 10^{3}$ |
| $p_{1}$ | $-6.1656 \times 10^{3}$ | $-3.9664 \times 10^{4}$ | $-2.3976 \times 10^{3}$ | $-409.4499$ |
| $p_{0}$ | $1.439 \times 10^{3}$ | $360.4347$ | $87.8580$ | $36.4711$ |

**Fig. S15. Polynomial fits the variations in $Q$ with $\kappa_{Si}$ for metasurfaces with different perturbations ($\Delta d$).**

The polynomial fits are defined by the equation $Q\left(\kappa_{Si}\right) = \sum_{n=0}^{N} p_n \kappa_{Si}^{n}$. For the fits we have $N_{50} = 13$, $N_{100} = 7$, $N_{200} = 4$, $N_{300} = 3$. The tabulated fit parameters are also provided.

**SI Note-22: Extracting $\beta_{eff}$ from thermal image data**

When the laser is switched off Eq. S4 assumes the form:

$$\frac{\partial \Delta T}{\partial t} = -\beta_{eff}\Delta T \quad \text{(S45)}$$

The solution to this equation is of the form:

$$T(t) + 273 = \varsigma e^{-\beta_{eff}\frac{t}{1000}} + \epsilon \quad \text{(S46)}$$

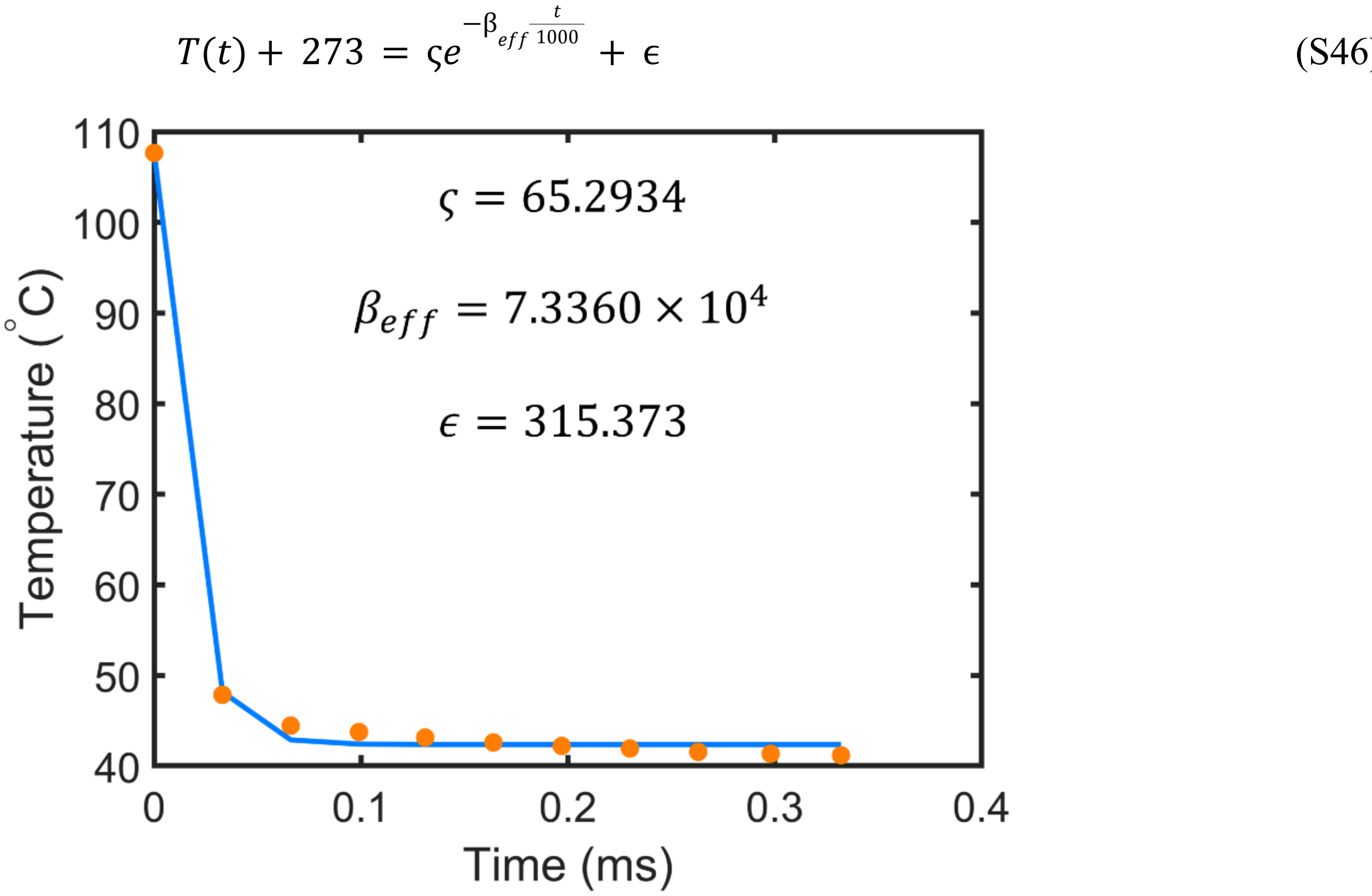

**Fig. S16. Extracting the heat loss rate ($\beta_{eff}$) by fitting Eq. S46 to thermal image data.** The extracted $\beta_{eff}$ is then used in Eq. S4 for all analysis.

**Supplementary References:**

1. J. Wang, B. Zhu, Z. Hao, F. Bo, X. Wang, F. Gao, Y. Li, G. Zhang, J. Xu, Thermo-optic effects in on-chip lithium niobate microdisk resonators. *Opt. Exp.* **24**, 19, 21869-21879 (2016).

2. O. Goldberg, R. Gherabli, J. Engelberg, J. Nijem, N. Mazurski, U. Levy, Silicon rich nitride huygens metasurfaces in the visible regime. *Adv. Opt. Mater.* **12**, 4, 2301612 (2024).

3. H. Jian, L. Yan. L. Senlin, L. Qing, Thermal conductivity of PECVD silicon-rich silicon nitride films measured with a $SiO_2/Si_xN_y$ bimaterial microbridge test structure. *J. Semicond.* **35**, 4, 046002 (2014).

4. H. Nejadriahi, Silicon-rich silicon nitride (SRN) for integrated photonics and thermo-optic applications. *Ph.D. Thesis in Electrical Engineering*, University of California San Diego (2022).

5. T. Lewi, H. A. Evans, N. A. Butakov, J. A. Schuller, Ultrawide thermo-optic tuning of PbTe meta-atoms. *Nano Lett.* **17**, 6, 3940-3945 (2017).

6. F. Pan, X. Li, A. C. Johnson, S. Dhuey, A. Saunders, M. Hu, J. P. Dixon, S. Dagli, S. Lau, T. Weng, C. Chen, J. Zeng, R. Apte, T. F. Heinz, F. Liu, Z. Deng, J. A. Dionne, Room-temperature valley-selective emission in Si-$MoSe_2$ heterostructures enabled by high-quality-factor chiroptical cavities. *Nat. Comm.* **17**, 20 (2026).